\documentclass[aps,prd,superscriptaddress,twocolumn]{revtex4}

\usepackage{graphicx}

\newcommand{\xleft}{{x_{\rm left}}}
\newcommand{\xright}{{x_{\rm right}}}
\newcommand{\xmid}{{x_{\rm mid}}}
\newcommand{\hmax}{{h_{\rm max}}}

\begin{document}

\title{Global structure of Choptuik's critical solution in scalar
field collapse} 

\author{Jos\'e M. Mart\'\i n-Garc\'\i a}
\email[]{J.M.Martin-Garcia@maths.soton.ac.uk} 
\affiliation{Faculty of Mathematical Studies, University of Southampton,
         Southampton SO17 1BJ, UK}

\author{Carsten Gundlach}
\email[]{C.Gundlach@maths.soton.ac.uk} 
\affiliation{Faculty of Mathematical Studies, University of Southampton,
         Southampton SO17 1BJ, UK}

\date{\today}


\begin{abstract}

At the threshold of black hole formation in the gravitational collapse
of a scalar field a naked singularity is formed through a universal
critical solution that is discretely self-similar. We study the global
spacetime structure of this solution. It is spherically symmetric,
discretely self-similar, regular at the center to the past of the
singularity, and regular at the past lightcone of the singularity.  At
the future lightcone of the singularity, which is also a Cauchy
horizon, the curvature is finite and continuous but not
differentiable. To the future of the Cauchy horizon the solution is
not unique, but depends on a free function (the null data coming out
of the naked singularity). There is a unique continuation with a
regular center (which is self-similar). All other self-similar
continuations have a central timelike singularity with negative mass.

\end{abstract}

\pacs{}

\maketitle




\section{Introduction} 


In general relativity, a black hole may be formed during the evolution
from asymptotically flat initial data where none is present. Consider
a one-parameter family of regular asymptotically flat initial data. It
is not difficult to find such families which form a black hole for
some range of the parameter (strong data) but disperse for another
range (weak data). The boundary between the two regimes is the black
hole threshold. In what is called type II critical collapse, the black
hole mass can be made arbitrarily small by adjusting the parameter $p$
of the initial data to its critical value $p_*$. Near the 
threshold, the final black hole mass $M$ then scales as
\begin{equation}
M\simeq C(p-p_*)^\gamma
\end{equation}
where $C$ is a constant. $C$ depends
on the family, but the transcendental number $\gamma$ is universal --
it depends on the type of matter but not on the family of initial
data.

Type II critical collapse was originally discovered by Choptuik in the
spherically symmetric massless scalar field \cite{Choptuik}, but has
since been found in many simple matter systems in spherical symmetry,
and also in axisymmetric gravitational waves \cite{AbrahamsEvans}. A
review of the field is \cite{critreview}.

Type II critical phenomena can be described in dynamical systems
terms: the phase space of the system has (at least) two attracting
fixed points, namely black holes and dispersion. The boundary between
the two basins of attraction, the critical surface, contains a
critical point: it is an attractor within the boundary surface, and a
repeller only in the one remaining direction. This means that it must
have precisely one unstable linear perturbation, with the property
that adding a bit of that perturbation with one sign leads to
collapse, while adding it with the opposite sign leads to
dispersion. In type II critical collapse the critical point is
either a discretely self-similar (DSS) or a continuously self-similar
(CSS) spacetime. 

Type II critical collapse is interesting, among other things, because
the maximum value of the curvature in a subcritical evolution, and the
maximal value of the curvature outside the black hole in a
supercritical evolution, both diverge as
\begin{equation}
(R_{abcd}R^{abcd})_{\rm max}\sim |p-p_*|^{-4\gamma}
\end{equation}
as the fine-tuning is improved. (This is basically the same result as
the black hole mass scaling, and similar results hold for any
curvature invariant.) From the dynamical systems picture it is clear
that the end point of type II critical collapse in the limit of
perfect tuning of $p$ to its critical value $p_*$ is not a ``zero mass
black hole'' but the critical solution itself. This solution has a
naked singularity. It is therefore interesting to examine the global
spacetime structure of the critical solution, and in particular its
Cauchy horizon. Here we do this for the spherically symmetric massless
scalar field, where the critical solution is DSS. We focus on this
system because CSS can be viewed as a limiting case of DSS, and
because the critical solution in the most interesting system in which
type II critical phenomena have been found, axisymmetric pure gravity,
is also DSS \cite{AbrahamsEvans}.

In Section \ref{section:kinematics} we discuss the global structure of
Choptuik's critical solution kinematically. Section
\ref{section:scalarfield} sets out the field equations for the real
massless scalar field in spherical symmetry, in coordinates adapted to
our problem, and describes the mathematical structure of the solution
at the Cauchy horizon of the singularity. Sections
\ref{section:numerics}, \ref{section:continuation} and
\ref{section:globalimages} show the results of our numerical
integration of the critical solution, and Section
\ref{section:conclusions} contains our conclusions. Some technical
details have been removed to appendices.


\section{Kinematical discussion}
\label{section:kinematics}


A spacetime is discretely self-similar (DSS) if there is a conformal
isometry $\Phi$ of the spacetime such that
\begin{equation}
\label{DSSgeometric}
\Phi_* g_{ab}=e^{-2\Delta}g_{ab}. 
\end{equation}
The value of the dimensionless ``logarithmic scale period'' $\Delta$
is a geometric property of the spacetime, independent of
coordinates. It is often useful to work in coordinates adapted to the
symmetry at hand. A generic self-similar and spherically symmetric
metric can be written as
\begin{equation}
\label{spherical_metric}
ds^2=e^{-2\tau}\left(A\,d\tau^2+2B\,d\tau\,dx+C\,dx^2+F^2\,d\Omega^2\right)
\end{equation}
where $d\Omega^2=d\theta^2+\sin^2\theta\,d\varphi^2$ is the metric on
the unit 2-sphere, and where $A$, $B$, $C$ and $F$ are functions of
$\tau$ and $x$. This metric is DSS if and only if they are periodic in
$\tau$ with period $\Delta$. It is continuously self-similar (CSS) if
these functions are completely independent of $\tau$.  We assume that
the signature is $(-,+,+,+)$, and that the metric is
non-degenerate. This leads to the inequality $AC-B^2<0$. We also assume
$F\ge 0$ for definiteness.

Any four-dimensional spacetime splits into a product of a
two-dimensional spacetime (the reduced manifold) and a round
two-sphere of area $4\pi r^2$. The area radius $r$ is a scalar in the
reduced manifold. Here the coordinates on the reduced manifold are
$\tau$ and $x$, and the area radius is given by $r=
e^{-\tau}F$. Geodesics in the reduced spacetime are radial (constant
$\theta$ and $\varphi$ geodesics in the full spacetime). The Hawking
mass $m$ is defined by $1-2m/r=(\nabla r)^2$. It is a scalar on both
the full and the reduced spacetime. From $m$ we define the two
dimensionless scalars $\mu= 2m/r$ and $a= (1-\mu)^{-1/2}$. It is easy
to show that a spherical surface where $\mu\ge1$ is a closed trapped
surface, and one where $\mu=1$ is an apparent horizon.
In a DSS spherical spacetime, $\mu$ and $a$ are periodic in $\tau$.

Radial null geodesics which are invariant under the symmetry
(\ref{DSSgeometric}) are called self-similarity horizons (SSH). They
are the key to determining the causal structure. All coordinate
systems of the form (\ref{spherical_metric}) are related by coordinate
transformations of the form
\begin{equation}
\label{dssgaugefreedom}
x'=\varphi(\tau,x),\qquad \tau'=\tau+\psi(\tau,x),
\end{equation}
where $\varphi$ and $\psi$ are periodic in $\tau$ with period
$\Delta$. We use this coordinate freedom to make all lines in the
reduced manifold where $F=0$ into lines of constant $x$. (These can be
either regular centers or central singularities.) We also make all SSHs
into lines of constant $x$ where $A=0$.

In order to discuss the global structure of the Choptuik spacetime and
its possible continuations we briefly review the kinematical results
of \cite{spherical_dss}. In a spherically symmetric DSS spacetime, two
kinds of singularities can be distinguished. From dimensional analysis
it can be seen that the Kretschmann scalar scales as $e^{4\tau}$ for
constant $x$, and
therefore the set $\tau=\infty$ is a central (because $r$ scales as
$e^{-\tau}$) curvature singularity. We call this the kinematical
singularity. Geometrically, this singularity is either a point or a
null line in the reduced spacetime. There are two types of
self-similarity horizons that in \cite{spherical_dss} 
we have called fans and splashes. The kinematical
singularity is null if there is at least one splash. Additional
central singularities can arise where $F=0$ for all $\tau$. (We call
these dynamical.) Because $\tau$ takes all values up to $\infty$ they
are connected to the kinematical singularity. There are at most two of
them, connected to the kinematical singularity at its
ends. Topologically, they are lines in the reduced manifold.

All known type II critical solutions in spherical gravitational
collapse can be defined by the properties of self-similarity
(CSS or DSS), analyticity at the past lightcone, and the requirement that
they have a single unstable perturbation mode. A generic spherically
symmetric DSS scalar field solution is singular either at the center
or the past lightcone. Imposing analyticity at both the center and the
past lightcone defines a nonlinear PDE boundary value-problem which
admits at most discrete solutions \cite{critscalar}. Only one such
solution has been found, and it empirically turns out to have only one
unstable mode, and to agree perfectly with the critical solution found
previously in collapse simulations by Choptuik \cite{Choptuik}.

\begin{figure}
\includegraphics[width=7cm]{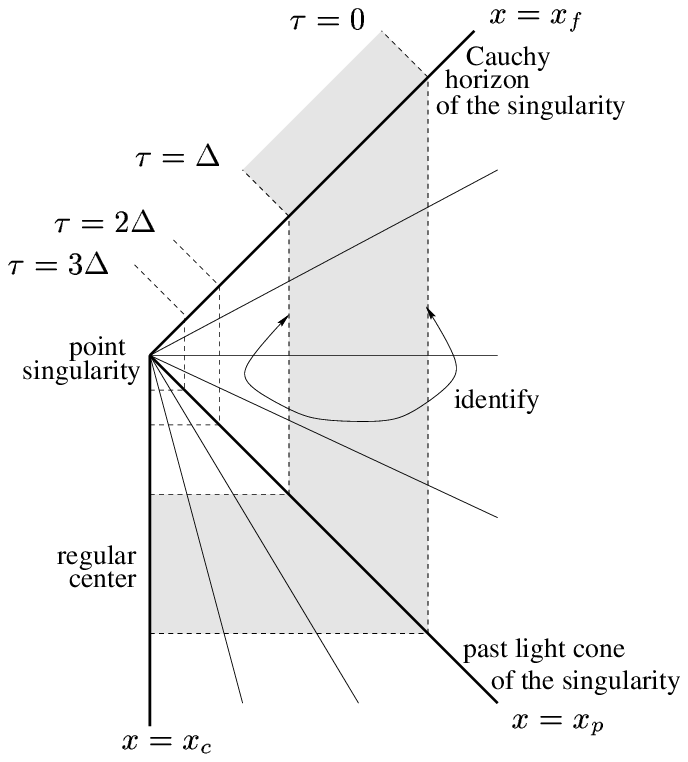}
\caption{ (Uncompactified) conformal diagram of the critical solution
up to the Cauchy horizon showing DSS-adapted coordinates. DSS lines
are shown continuous. Lines of constant $\tau$ are shown dashed (we
have assumed $e^\Delta=2$ while in reality $e^\Delta\simeq 31$). Note
that the numerical domain is $0\le \tau<\Delta$ (shaded), with
periodic boundary conditions. We have illustrated the three coordinate
patches we use for numerical work: in the past patch between $x_c$ and
$x_p$ $\tau$-lines are spacelike. In the outer patch between $x_p$ and
$x_f$ $\tau$-lines are timelike. In the future patch beyond $x_f$ they
are null. The three patches together cover the entire spacetime
without overlapping, and the coordinates $x$ and $\tau$ are continuous
at the interfaces.}
\label{figure:preCH}
\end{figure}

The global structure of the Choptuik solution up to the future
light cone of the kinematical singularity (which is a Cauchy horizon)
is sketched in Fig.~\ref{figure:preCH}, together with the $x$ and
$\tau$-lines of the three coordinate patches that we shall use in the
numerical calculations.  This structure is the same as for all other
known type II critical solutions in spherical symmetry. These
solutions have a regular center $x=x_c$ in the past. As $x$ increases,
the $x$ lines are at first timelike, so that $A<0$. They become null
at the fan $x=x_p$ where $A=0$, $\partial A/\partial x>0$ and
$B>0$. As $x$ increases further, they are spacelike, so that $A>0$
spacelike. Somewhere in the spacelike region $B$ changes sign. The
$x$-lines become null again at the second fan $x=x_f$, where $A=0$,
$\partial A/\partial x<0$ and $B<0$. Approaching the kinematical
singularity $\tau=\infty$ from the range of ``angles'' $x_c\le x \le
x_f$ it is a single point. $x_p$ and $x_f$ are its past and future
lightcones.

We demonstrate below {\em numerically} 
that the curvature at the Cauchy horizon $x=x_f$
is finite in Choptuik's scalar field critical solution. Furthermore
all geodesics cross it in finite affine parameter. The spacetime can
therefore be extended beyond, but this continuation is not
unique. Mathematically speaking we shall see that the solution is not
analytic at the Cauchy horizon in the limit coming from the past, and
so there is no preferred analytic continuation to the future. (The
curvature is only $C^0$ from the past.) The family of continuations
in which the curvature is $C^0$ across the Cauchy horizon and which
are DSS is parameterized by one free periodic function $\hat
U_\epsilon(\tau)$. Physically speaking this function can be
interpreted as data on the naked singularity which determine the
continuation, in addition to the null data on the Cauchy horizon.

We now discuss the possible continuations that are allowed
kinematically. 

\begin{figure}
\includegraphics[height=8cm]{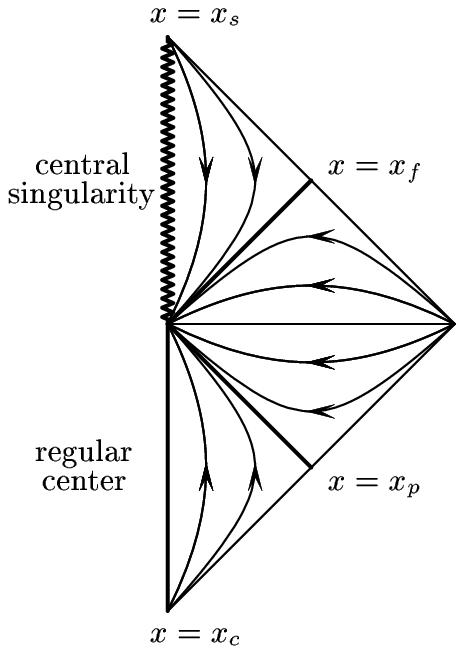}
\caption{An extension with a timelike central singularity. Instead of
the central singularity the solution could also have a regular
center. This diagram has two fans at $x_p$ and $x_f$.}
\label{figure:timelikesingular}
\end{figure}

In the simplest case, the $x$-lines to the future of the Cauchy
horizon are all timelike ($A<0$) until a timelike singularity $F=0$ is
reached. In our classification this is a dynamical singularity, while
the kinematical singularity is a single point. This conformal
structure is shown in Fig.~\ref{figure:timelikesingular}. If $m\sim
r^3$, or equivalently $\mu\sim F^2$ as $x\to x_r$, the conformal
structure is the same, but with the singularity replaced by a regular
center.

\begin{figure}
\includegraphics[height=8cm]{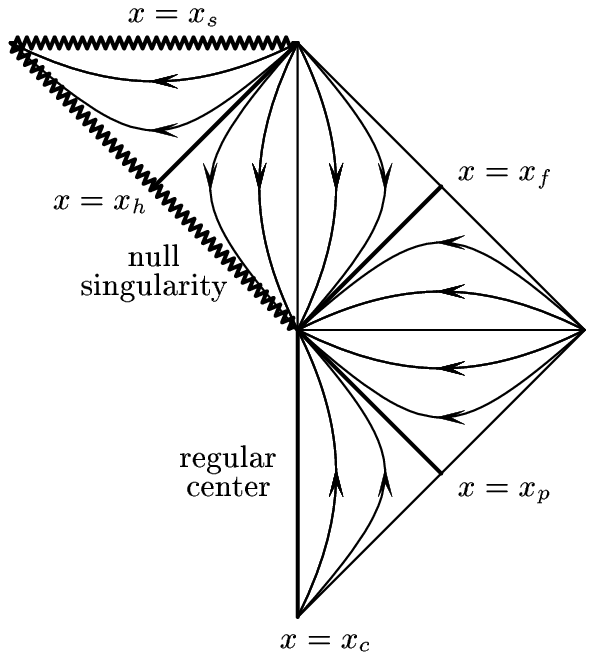}
\caption{A hypothetical extension with a null singularity. This
diagram has two fans at $x_p$ and $x_f$ and a splash at $x_h$. Note
that this extension is {\em not} realized as a DSS continuation of the
Choptuik solution.  }
\label{figure:spacelikesingular}
\end{figure}

Nolan \cite{Nolan} has drawn spacetime diagrams for spherically
symmetric CSS spacetimes in which the point singularity at the origin
of the Cauchy horizon is only the starting point of an ingoing central
null singularity. In our classification this is an extended
kinematical singularity, which requires the existence of at least one
splash, that is a line where $A=0$
again. Fig.~\ref{figure:spacelikesingular} shows the simplest generic
possibility, in which the splash is followed by a spacelike dynamical
singularity, which covers part of the naked null singularity. This
spacetime structure can actually be realized in spherical CSS scalar
field solutions \cite{OriPiran}. The same, and more exotic structures,
can also be realized in spherical CSS perfect fluid solutions
\cite{fluidcss}.

Contrary to our expectations, we have found that none of these exotic
possibilities are realized as spherical DSS continuations of the
Choptuik solution.  There is a unique DSS continuation with a regular
center, and all other DSS continuations have a timelike
singularity. In hindsight, the reason appears to be that the null data
on the Cauchy horizon are extremely weak. With stronger data (not
associated with the Choptuik spacetime) we find different kinds of
continuations.

\section{Coordinates and field equations}
\label{section:scalarfield}


In spherical symmetry, there are four algebraically independent
Einstein equations, which can be taken to be $G_{\tau\tau}$, $G_{\tau
x}$, $G_{xx}$ and $G_{\theta\theta}$. The fourth of these is a
combination of derivatives of the first three and can therefore be
disregarded. The first three equations contain first and second
derivatives of $F$, but only first derivatives of $A$, $B$, and $C$.
In the following we assume that $F$ is a given function of $x$ and
$\tau$.  The Einstein equations then determine three independent
linear combinations of the six first derivatives of $A$, $B$ and
$C$. A non-degenerate basis of this three-dimensional space is
$\mu_{,x}$, $\mu_{,\tau}$ and $\nabla_\mu\nabla^\mu r$. (With
$F(\tau,x)$ given, this last term contains only first derivatives of
$A$, $B$ and $C$.)

Here we investigate massless real scalar field matter. The scalar
field $\phi$ obeys the wave equation
\begin{equation}
\nabla_a\nabla^a\phi=0,
\end{equation}
and the Einstein equations can be written as
\begin{equation}
\label{Ricci}
R_{ab}=8\pi G\,\nabla_a\phi \nabla_b\phi.
\end{equation}
The wave equation, when written as a second order PDE, will in general
contain both first derivatives of all four metric
coefficients. However, by writing the wave equation in a geometrically
defined first-order form, all metric derivatives except one can be
eliminated. Define two null derivative operators $\nabla_u$ and
$\nabla_v$ with the usual convention that both point towards the
future and that $\nabla_u$ points inwards while $\nabla_v$ points
outwards. Normalize them by imposing that $\nabla_ur=-r$ and
$\nabla_vr=r$. We define the null derivatives of the scalar field as
$U=\sqrt{2\pi G} \, \nabla_u\phi$ and $V=\sqrt{2\pi G}\, \nabla_v\phi$.
The massless wave equation in spherical symmetry can then be written as
\begin{eqnarray}
\nabla_vU-V+PU&=&0,\\
\nabla_uV+U-PV&=&0,
\end{eqnarray}
with the scalar $P\equiv r\nabla^2r /(\nabla r)^2 = a^2 - 1$.
Using the Einstein equations, the curvature can be given in terms of
$U$ and $V$:
\begin{equation}
R_{uu}=4U^2, \quad R_{vv}=4V^2, \quad R_{uv}=4UV,
\end{equation}
(other components of the Ricci tensor vanish) and
\begin{equation}
R= \frac{-4}{a^2 r^2}UV .
\end{equation}

The most general form of the scalar field compatible with DSS is 
\begin{equation}
\label{generalphi}
\phi(\tau,x)=\psi(\tau,x)+\kappa \tau
\end{equation}
where $\psi(\tau+\Delta,x)=\psi(\tau,x)$ and $\kappa$ is a global
constant. In the Choptuik solution, $\kappa=0$ empirically, so that
$\phi$ is periodic in $\tau$. This means that $U$ and $V$ have zero
average in $\tau$. Moreover, $U$ and $V$ in the Choptuik solution obey
$U(\tau+\Delta/2,x)=-U(\tau,x)$ and so for $V$. (This of course
implies zero average.) As a consequence
$\mu(\tau+\Delta/2,x)=\mu(\tau,x)$ and so for other suitable metric
fields. We shall assume these extra symmetries in our numerical work,
but all our analytical expressions remain valid if these symmetries
are dropped.

We now describe three coordinate patches that cover the critical
solution. We demand that both the past and the future lightcones 
of the singularity occur
at lines of constant $x$. This makes it easier to impose regularity at
the center and past lightcone, and to investigate the behavior at the
future lightcone (Cauchy horizon). It also allows us to match the
coordinate patches without overlap. Subject to these requirements, we
have tried to make the field equations in each patch as simple as
possible. Based on the three patches, it is straightforward to
construct a single smooth coordinate system covering the whole
spacetime (see Section \ref{section:globalimages}), but using it from
the beginning would unnecessarily complicate our numerical work.  We
summarize a number of coordinate systems for spherically symmetric CSS
and DSS spacetimes, and their advantages and disadvantages, in the
Appendix.


\subsection{Past patch}


On the past patch, which extends from the regular center to the past
lightcone, we write the metric in terms of two free functions
$f(\tau,x)$ and $a(\tau,x)$ as
\begin{equation}
\label{pastpatch}
A=a^2(x^2-f^2), \qquad
B=-xa^2, \qquad
C=a^2, \qquad
F=x,
\end{equation}
for $x\ge x_c\equiv 0$. Here $a=(1-2m/r)^{-1/2}$ is the scalar we
defined above. These coordinates can be derived from the
Schwarzschild-type metric
\begin{equation}
ds^2=-\alpha^2\,dt^2+a^2\,dr^2+r^2\,d\Omega^2
\end{equation}
through the coordinate change $x=r/(-t)$ and $\tau=-\ln(-t)$, and
defining $f=\alpha/a$. The remaining gauge freedom of time
relabelling is fixed by imposing that the past lightcone is on a
constant $x=x_p$ line. Note that we do not follow the convention of
\cite{Choptuik} that $\alpha=1$ at the center, which in
\cite{critscalar} forced us to introduce an additional free function
of time in the definition of $x$. The Einstein equations are
\begin{eqnarray} \label{pastmetriceqs}
f_{,x}&=&{(a^2-1)f\over x},\\
(a^{-2})_{,x}&=&{1-(1+U^2+V^2)a^{-2}\over x},\\
(a^{-2})_{,\tau}&=&\left[{(f+x)U^2-(f-x)V^2\over x} +1\right]a^{-2}-1, 
\end{eqnarray}
and the matter equations are
\begin{eqnarray} \label{pastmattereqs}
U_{,x}&=&{f[(1-a^2)U+V]-{xU_{,\tau}}\over x(f+x)},
\\
V_{,x}&=&{f[(1-a^2)V+U]+{xV_{,\tau}}\over x(f-x)},
\end{eqnarray}

At the regular center $x=0$ we impose elementary flatness, that is the
absence of a conical singularity. In order to do this, we define
$\Pi=(V+U)/(2x)$ and $\Psi=(V-U)/(2x^2)$ and impose that both are
regular even functions of $x$ at $x=0$. At the past lightcone we have
$f-x=0$, which by our gauge choice means that $f=x_p$ there.  We also
impose the physical regularity condition
\begin{equation}
\label{past_patch_regularity}
V_{,\tau}-(1-a^2)V-U=0
\end{equation}
on the past lightcone. The conditions of DSS, regularity at the
center and regularity at the past lightcone select a solution. The
equations on the past patch are form-invariant under the linear
coordinate transformation $x\to cx$, $f\to cf$, and $a$, $U$ and $V$
unchanged. In the numerical results presented here we have set $x_p=1$
on the past patch. Note that the regularity condition 
(\ref{past_patch_regularity}) is coordinate-independent, as $U$, $V$ and
$a$ and $r$ are all scalars and $\tau=-\ln r$ on the lightcone.


\subsection{Outer patch}


On the outer patch, which extends from the past to the future lightcone,
we write the metric in terms of $a(\tau,x)$, $b(\tau,x)$, and the
auxiliary function $\xi(\tau)$ as
\begin{equation}
\label{outerpatch}
A=a^2(1-b^2), \qquad
B=a\, b\, \xi, \qquad
C=-\xi^2, \qquad F=1,
\end{equation}
where $\xi>0$ is a function of $\tau$ only. $a>0$ is the scalar
defined above. We fix the remaining gauge freedom by imposing that the
past lightcone $b=-1$ occurs at $x=x_p$, and the future lightcone
$b=1$ at $x=x_f$. Putting both lightcones on a line of constant $x$
requires $\xi(\tau)$ to be non-constant. There is no outer coordinate
patch that does not have at least one function like $\xi(\tau)$. 
As $\tau=-\ln r$ everywhere on the outer patch, $\tau$ is continuous
between the past and outer patches.

The metric equations are
\begin{eqnarray}
\label{outer_first}
\frac{b_{,x}}{\xi}&=&
-\frac{-3+a^2+U^2(1-b)+V^2(1+b)}{2a}-\frac{\xi'}{a\xi}\; , \\
\label{outer_second}
\frac{a_{,x}}{\xi}&=&-\frac{U^2-V^2}{2}, \\
\label{outer_a}
\left(a^{-2}\right)_{,\tau}&=& 
\left[1+U^2(1-b)+V^2(1+b)\right]\, a^{-2}-1,
\end{eqnarray}
and the matter equations are
\begin{eqnarray}
\frac{U_{,x}}{\xi}=\frac{(1-a^2)U+V+U_{,\tau}}{a(1-b)}, 
\label{outer_U} \\
\frac{V_{,x}}{\xi}=-\frac{(1-a^2)V+U+V_{,\tau}}{a(1+b)}.
\label{outer_V}
\end{eqnarray}
Note that the $a_{,\tau}$ constraint equation is again linear in 
$a^{-2}$. The equations on the outer patch are form-invariant under the
linear coordinate transformation $x\to cx+d$, $\xi\to \xi/c$, and $a$, 
$b$, $U$ and $V$ unchanged. In the numerical results presented here we
have set $x_p=-1$ and $x_f=1$ on the outer patch.


\subsection{Singular behavior at the Cauchy horizon}
\label{section:singularCH}


We shall see that at the Cauchy horizon the solution is mildly
singular. Naive finite differencing breaks down there. Instead
we expand the generic solution around the Cauchy horizon in terms of
two free periodic functions $V_0$ and $\hat U_\epsilon$, and match
this expansion to the numerical evolution at a small finite distance
to the past of the Cauchy horizon. Before describing the full
expansion, we focus on the origin of the singular behavior.

Equation (\ref{outer_U}) becomes singular at the future lightcone
because the denominator of the right hand side vanishes there. In
contrast to the past lightcone, we do not have any freedom left to
enforce the vanishing of the numerator as well. Therefore we expect the
solution to have some kind of singularity at $x_f$. We rewrite
(\ref{outer_U})
\begin{equation}
D U_{,x} - U_{,\tau} - (1-a^2) U = V ,
\label{U_singular_eq}
\end{equation}
where we have defined the metric function
\begin{equation}
\label{Ddef}
D \equiv \frac{a(1-b)}{\xi} .
\end{equation}
$D$ is positive on the outer patch and vanishes on the future light
cone. The characteristics $x(\tau)$ of Eq. (\ref{U_singular_eq}) are
given by
\begin{equation}
\label{Ucharacteristics}
\frac{dx(\tau)}{d\tau}=-D\left(\tau,x(\tau)\right) .
\end{equation}
Recall that we impose the gauge condition that the CH is at $x=x_f$,
or $b(\tau,x_f)=1$. We define the shorthand $y=x-x_f$.

We now make one fundamental assumption, namely that the spacetime
admits regular null data on the Cauchy horizon $y=0$.  This assumption
is clearly necessary if we want to continue the spacetime through the
CH, but here we make it simply because we have not been able to find a
more general ansatz. We shall
see later that it is sufficiently general to be matched to
the critical solution that we have obtained numerically on the past
patch.

We therefore assume that $V$ is continuous, or
\begin{equation}
V(\tau,x)=V_0(\tau)+o(y^0).
\end{equation}
By substituting this into Eq. (\ref{outer_a}) in the limit $b=1$ we
find that $a$ is also continuous and $a_0(\tau)$ obeys
\begin{equation}
\label{a0eqn}
(a_0^{-2})'-(1+2V_0^2)\,a_0^{-2}+1=0.
\end{equation}
Because we impose periodic boundary conditions in $\tau$ this ODE has a
unique solution. The physical significance of this is that the null
data $V_0$ determine the geometry of the CH. Similarly, from
(\ref{U_singular_eq}) in the limit $D=0$ 
we find that $U$ must be continuous, and
$U_0(\tau)$ is the unique periodic solution of
\begin{equation}
\label{U0eqn}
U_0'+(1-a_0^2)U_0+V_0=0.
\end{equation}
This condition follows from the assumption of DSS.  Finally, from
(\ref{Ddef}), (\ref{outer_first}) and (\ref{outer_second}) 
we find that $D$ is once
differentiable, so that $D(\tau,x)=y\,D_1(\tau)+o(y)$, and $D_1$ is
given by
\begin{equation}
D_1={\xi'\over \xi}+{1\over 2}\left(-3+a_0^2+2V_0^2\right).
\end{equation}

At this point we introduce more shorthand notation. If $f(\tau)$ is
any periodic function (with period $\Delta$), let $\bar f$ be its
average value, and let $\tilde f(\tau)=f(\tau)-\bar f$ be its
oscillatory part. Let $\int \tilde f$ be the definite integral
$\int_{\tau_0}^\tau \tilde f(\tau')\,d\tau'$ where $\tau_0$ is
chosen so that $\int \tilde f$ has vanishing average.

We can now integrate (\ref{Ucharacteristics}) for the
$U$-characteristics to leading order and obtain
\begin{equation}
\log |y(\tau)| + \bar D_1 \tau + 
\int \tilde D_1 + o(y^0) = {\rm const.}
\label{characteristics}
\end{equation}
We see that on a characteristic $\tau\to-\infty$ as $|y|\to
0$. Because $U(\tau,x)$ is periodic in $\tau$ with period $\Delta$, an
infinite number of oscillations in $y$ at constant $\tau$ pile up at
the Cauchy horizon $y=0$. We can solve (\ref{U_singular_eq}) to
leading order by the method of characteristics. The general solution is
\begin{equation}
\label{Uepsilon}
U(\tau,y)= U_0(\tau)+|y|^\epsilon \check U_\epsilon(\tau)\, \hat
U_\epsilon(\hat\tau) + o(|y|^\epsilon)
\end{equation}
where $\hat U_\epsilon(\hat\tau)$ is an arbitrary periodic function
with period $\Delta$ and 
\begin{eqnarray}
\epsilon&\equiv&{1-\overline{a_0^2} \over \bar D_1} , \\
\check U_\epsilon(\tau)&\equiv&
\exp\left(\int\tilde {a_0^2}+\epsilon\int\tilde D_1\right), \\
\hat\tau&\equiv&\tau+H(\tau)+K\ln |y|, \\
H(\tau)&\equiv&\frac{1}{\bar D_1}\int \tilde D_1, \\
K&\equiv&\frac{1}{\bar D_1}. 
\end{eqnarray}

Rewriting (\ref{a0eqn}) as
\begin{equation}
a_0^2-(1+2V_0^2)=2(\ln a_0)',
\end{equation}
we see that $\overline{a_0^2}=1+2\overline{V_0^2}$,
and so we can express $\epsilon$ and $K$ as
\begin{equation}
\epsilon={2\overline{V_0^2} \over 1-2\overline{V_0^2}}, 
\qquad K=-(1+\epsilon).
\end{equation}
Our initial assumption that $U$ and $V$ are continuous at the light
cone is therefore justified either if $0<\overline{V_0^2}<1/2$, so
that $\epsilon>0$, or if $\hat U_\epsilon(\tau)=0$. We shall show
numerically that $\epsilon$ is small but positive on the CH. In this
case $U$ and $V$ are just $C^0$, and the scalar field is therefore
$C^1$. In spherical symmetry the Riemann tensor is determined
completely by the Ricci tensor, which in turn is quadratic in the
partial derivatives of $\phi$, see (\ref{Ricci}). The curvature is
therefore quadratic in $U$ and $V$ and so is $C^0$.

A similar analysis, with $U$ and $V$ interchanged, applies to the past
lightcone. In the notation we have introduced here we can describe the
past lightcone by saying that $\epsilon<0$ there (because the null
data $U_0$ on the past lightcone are large, $\overline{U_0^2}>1/2$) 
but that the free
coefficient $\hat V_\epsilon$ vanishes identically (because we have
imposed analyticity as a boundary condition).


\subsection{Expansion near the Cauchy horizon}
\label{section:expansion}


We cannot apply Fuchsian techniques to our system of equations because
they require the {\em simultaneous} vanishing on $y=0$ of the
coefficients of $U_{,x}$ and $U_{,\tau}$ in equation
(\ref{U_singular_eq}). 

However, the form (\ref{Uepsilon}) of the leading terms in $U$
suggests that the full non-linear solution can be written as a regular
part, containing only integer powers of $y$, plus a singular part
which contains powers of $|y|^\epsilon$.  We can in fact construct a
formal solution near the Cauchy horizon in the form of an asymptotic
double series: 
\begin{equation}
f(\tau,x)=\sum_{n=1}^\infty y^n f_n(\tau)
+\sum_{n=1}^\infty \sum_{k=0}^{k_{\rm max}(n)} |y|^{n+k\epsilon}
f_{n+k\epsilon}(\tau,x), 
\end{equation}
where $f$ stands for $U$, $V$, $a$ and $b$, and
\begin{equation}
f_{n+k\epsilon}(\tau,x)=
\sum_{i=1}^{i_{\rm max}(n,k)} \check f_{n+k\epsilon}^{(i)}(\tau)
\hat f_{n+k\epsilon}^{(i)}(\hat\tau).
\end{equation}
Here $\epsilon$ and $\hat\tau$ are the quantities defined above in
terms of $V_0$. The expansion depends on the two free periodic
functions $V_0(\tau)$ and $\hat U_\epsilon(\hat\tau)$. By function
counting we can therefore match any DSS solution to this expansion.

The first non-integer term appears in each variable at the following
orders:
\begin{eqnarray}
U(\tau,x)&=&U_0(\tau)+|y|^\epsilon \check U_\epsilon(\tau)\, 
\hat U_\epsilon(\hat\tau) \nonumber \\ &&
\qquad +y\,U_1(\tau)+O\left(|y|^{1+\epsilon}\right), 
\label{expansion0} \\
V(\tau,x)&=&V_0(\tau)+y\,V_1(\tau)+O\left(|y|^{1+\epsilon}\right),
\label{expansion1} \\
a(\tau,x)&=&a_0(\tau)+y\,a_1(\tau)+O\left(|y|^{1+\epsilon}\right), 
\label{expansion2} \\
b(\tau,x)&=&1+y\,b_1(\tau)+y^2\,b_2(\tau)+O\left(|y|^{2+\epsilon}\right)
. \label{expansion3}
\end{eqnarray}
In the previous section we obtained the coefficients of expansions
(\ref{expansion0}--\ref{expansion3}) up to $O(|y|^\epsilon)$.
Stopping there, the first order we neglect is $O(y)$. This truncation
already depends on both free functions $V_0$ and $\hat U_\epsilon$ and
shows the singular behavior. It is also a sensible truncation
numerically because $\epsilon$ turns out to be very small in the
Choptuik solution.  Going further, for the same reason there would be
no point in including $O(y)$ terms without also including all
$O(|y|^{1+k\epsilon})$ terms. It turns out that we need to go to
$O(|y|^{1+3\epsilon})$. We have used the expansion to that order to
check convergence. The expressions are given in Appendix
\ref{singular_expansion}.


\subsection{Future patch} 


Our analysis of the possible continuations of the critical solution in
Section~\ref{section:kinematics} has shown that we can cover the entire
future
of the Cauchy horizon in a single patch if we make $\tau$ an ingoing
null coordinate. This means setting $C=0$ and $B<0$. In order to put
the center $r=0$ at a known coordinate location, we also set
$F=-x$. We choose $x<0$ here so that $x$ increases as we extend the
spacetime away from the Cauchy horizon. We parameterize this metric in
terms of the scalar $a$ and a coefficient $f$ (not the same as $f$
in the past patch):
\begin{equation}
A=-4a^2 f(f+x), \quad
B=2a^2 f, \quad
C=0, \quad
F=-x.
\end{equation}
Regularity of the metric requires $a>0$ and $f>0$.
The field equations are
\begin{eqnarray}
\label{futureeq1}
f_{,x}&=&{(a^2-1)f\over x}, \\
(a^{-2})_{,x}&=&{1-(1+2U^2)a^{-2}\over x}, \\
\label{a_constraint}
(a^{-2})_{,\tau}&=&\left[-{2fV^2-2(f+x)U^2\over x}+1\right]a^{-2}-1, \\
U_{,x}&=&{f[(1-a^2)U+V]-xU_{,\tau}\over x(f+x)}, \\
V_{,x}&=&{U+(1-a^2)V\over x}.
\label{futureeq5}
\end{eqnarray}
At the Cauchy horizon $x=x_f<0$ we impose the coordinate condition
$f+x=0$. The equations are invariant under $x\to cx$, $f\to cf$. We
use this to set $x_f=-1$ on this patch. As a consequence $\tau=-\ln r$
on the lightcone, and so $\tau$ is continuous between the outer and
future patches.

We find an asymptotic expansion around the Cauchy horizon in terms of
two free functions $V_0$ and $\hat U_\epsilon$. $V_0$ is given by the
null data on the Cauchy horizon, and so is the same as on the outer
patch. $\epsilon$ is therefore the same on both sides. $U_0$ obeys the
same ODE, (\ref{U0eqn}), as on the outer patch, and so is the same
function. There is no need, however, to make $\hat U_\epsilon$ the
same on both sides, as we would not gain any differentiability by
doing so. Instead we consider $\hat U_\epsilon$ as free ``data on the
naked singularity'', and we shall find experimentally how the global
structure of the spacetime is influenced by this choice.


\section{Numerical construction of the Choptuik spacetime up to the
Cauchy horizon}
\label{section:numerics}


We have carried out a brand new numerical computation of the Choptuik
spacetime, improving the precision of our previous calculations 
\cite{critscalar} by roughly four orders of magnitude.
This was mainly needed to assert without doubt that $\epsilon$ is
different from zero, even though it is extremely small.

Essentially, our new scheme uses shooting methods on the $x$ axis,
instead of relaxation methods. This allows us to improve the treatment
of the regular singular points of the equations (the center and the
lightcones) by using Taylor expansions. We still work with
pseudo-spectral Fourier techniques in $\tau$ because the solution is
periodic.  We shall see however that the particular structure of the
Fourier transform of the Choptuik spacetime poses an unexpected
problem when combining Taylor and Fourier expansions.

In this Section we explain the numerical scheme and present the results
for the first two patches, with special emphasis on convergence
properties and error analysis.


\subsection{Numerics}


\subsubsection{Pseudo-spectral decomposition}


For definiteness, let us suppose we work on the past patch.  Following
\cite{critscalar} we discretize our $\Delta$-periodic fields
$Z(\tau,x)$, where $Z$ stands for any of the set $\{a,f,U,V\}$,
using $N$ equidistant points in one period:
\begin{equation}
Z_n(x)\equiv Z\left(\frac{n}{N}\Delta,x\right)= 
\sum_{k=0}^{N-1} \hat{Z}_k(x) e^\frac{2\pi i k n}{N} 
\end{equation}
for $n=0,...,N-1$. In this way we transform our 1+1 PDE problem for 
$Z(\tau,x)$ into an ODE problem for the modes $\hat{Z}_k(x)$.
The essential idea of pseudo-spectral methods is to
carry out algebraic operations pointwise on the $Z_n$ and 
$\tau$-differentiation/integration on the $\hat{Z}_k$, switching
from one to the other with a Fast Fourier Transform algorithm.

The main drawback of the method is the aliasing problem: pointwise
products of fields (nonlinearities) generate high frequency modes
which cannot be sampled with only $N$ points. We partially solve that
problem by doubling the $\hat{Z}_k$ (padding with zeros) before going
to the $Z_n$, carrying out the necessary algebraic operations on the
doubled $Z_n$ and going back to the $\hat{Z}_k$, then halving the
$\hat{Z}_k$ and thus throwing away high frequency noise.  We have
tried other possibilities, such as padding with $3N$ or $7N$ instead
of $N$ zero Fourier components, or extrapolating the Fourier
coefficients using the observed fact that high frequency modes have a
simple exponential dependence on frequency (see below), but the
results are not improved. Aliasing can only be reduced by going to
higher $N$. From a numerical point of view, we are only safe from
aliasing when the amplitude of the modes we are cutting off is below
machine precision.

Because all our fields are real $\hat{Z}_k^*=\hat{Z}_{N-k}$.
Furthermore, the metric fields $a$ and $f$ are even [in the sense 
$a(\tau+\Delta/2,x)=a(\tau,x)$] and therefore their Fourier transform
only contains even $k$ modes, while the matter fields $U$ and $V$ are
odd [in the sense $U(\tau+\Delta/2,x)=-U(\tau,x)$] and therefore their 
Fourier transform only contains odd $k$ modes.
Taking this symmetry into account an even or odd field $Z$ sampled
with $N$ points per period is encoded by $N/4$ independent non-zero
complex modes. As we have 4 independent variables $a,f,U,V$,
the ODE system we solve comprises $N$ complex or $2N$ real variables.
In our calculations we have used $N=32,64,128,256$ and $512$.
Previous investigations used $N=64$. See Appendix C of Ref.
\cite{critscalar} for a complete discussion of our Fourier
pseudo-spectral method.


\subsubsection{Shooting to fitting points}


We cannot cross the lightcones during the integration of the ODE in
the $x$-axis because they are regular singular points of the equations.
Therefore we perform consecutive shooting calculations on the past,
outer and future patches, in this order because we 
need information from the first to shoot the second and from the second
to shoot the third. The issue of error propagation becomes very
important.

Again, we describe the past patch for definiteness.  Given the
equations (\ref{pastmetriceqs}--\ref{pastmattereqs}) and free data
$f(\tau, x_c)\equiv f_c(\tau)$, $\Psi(\tau, x_c) \equiv \Psi_c(\tau)$
at the center, we calculate the solution at $x_{\rm left}$ slightly
larger than $x_c$ using a second-order power expansion [leaving errors
of order $(\xleft-x_c)^4$].  From these data we integrate the ODE system
forward in $x$ using finite differencing, up to $x_{\rm mid}$. In the
same way, given free data $U(\tau, x_p)\equiv U_p(\tau)$ and the gauge
condition $f(\tau,x_p)=x_p$, we calculate the solution at $x_{\rm
right}$ slightly smaller than $x_p$ [this time with errors of order
$(x_p-\xright)^3$] and integrate backward in $x$ up to the same $x_{\rm
mid}$. Finally we use Newton's method to look for the free data which
brings the mismatch between both integrated solutions at $x_{\rm
mid}$ down to a minimum, typically of order $10^{-13}$.
(This is the machine precision of
$10^{-16}$ reduced by a factor $10$ due to the calculations at each
step and a factor $100$ from the ODE-integration along $\approx 10^4$
points.)

We use a grid that becomes logarithmic near the regular singular
points, with maximum stepsize $\hmax$ at some intermediate position.
That grid is constructed using the transformation
\begin{equation}
x= \frac{x_c+x_p e^z}{1+e^z}
\end{equation}
from a grid of equidistant points in $z$ between the values 
$z_{\rm left}$ and $z_{\rm right}$ corresponding to $x_{\rm left}$ and
$x_{\rm right}$ respectively. Near the two endpoints we have
$x-x_c\simeq (x_p-x_c) e^z$ and $x_p-x\simeq (x_p-x_c) e^{-z}$. 
We integrate on a fixed grid in 
$x$ rather than using a variable stepsize method in order to check for
convergence with $\hmax$. This gives us a good estimate of the
underlying discretization error.

Concerning the ODE-integrator, we have tried several Runge-Kutta
methods, both explicit and implicit (Gauss-Legendre), with different
convergence orders \cite{Iserles}.  In general, implicit methods are
better suited to our problem than explicit methods, particularly for
high $N$, because high frequencies make the problem stiffer.  We
choose an IRK2 method (implicit, 2 stages, order 4), which is a
compromise between the accuracy of a high order method and the clarity
of convergence of a low order method. Our implicit schemes are
implemented by iteration until $l_2$-differences between successive
iterations converge below $10^{-15}$. We cannot get closer to actual
machine precision ($10^{-16}$) due to the accumulation of roundoff
error. An implicit step typically takes 5 to 15 iterations to
converge.


\subsection{Past patch}

\begin{figure}
\includegraphics[width=\columnwidth]{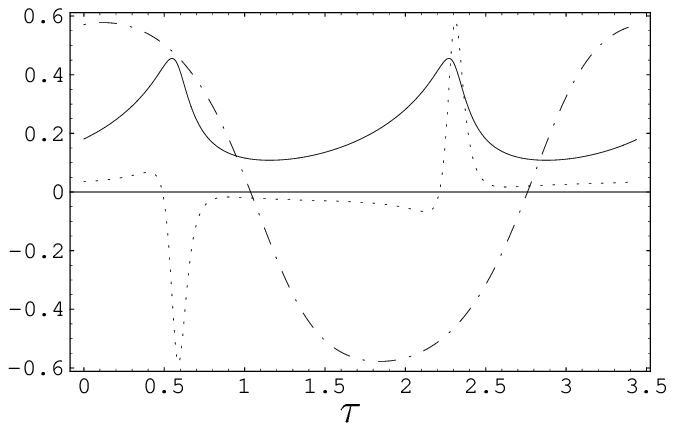}
\caption{\label{freedata}
Best free data on the singular points: $f_c(\tau)$ (continuous line),
$\Psi_c(\tau)/300$ (dotted line) and $U_p(\tau)$ (dash-dot line).
}
\end{figure}
\begin{figure*}
\includegraphics[width=15cm]{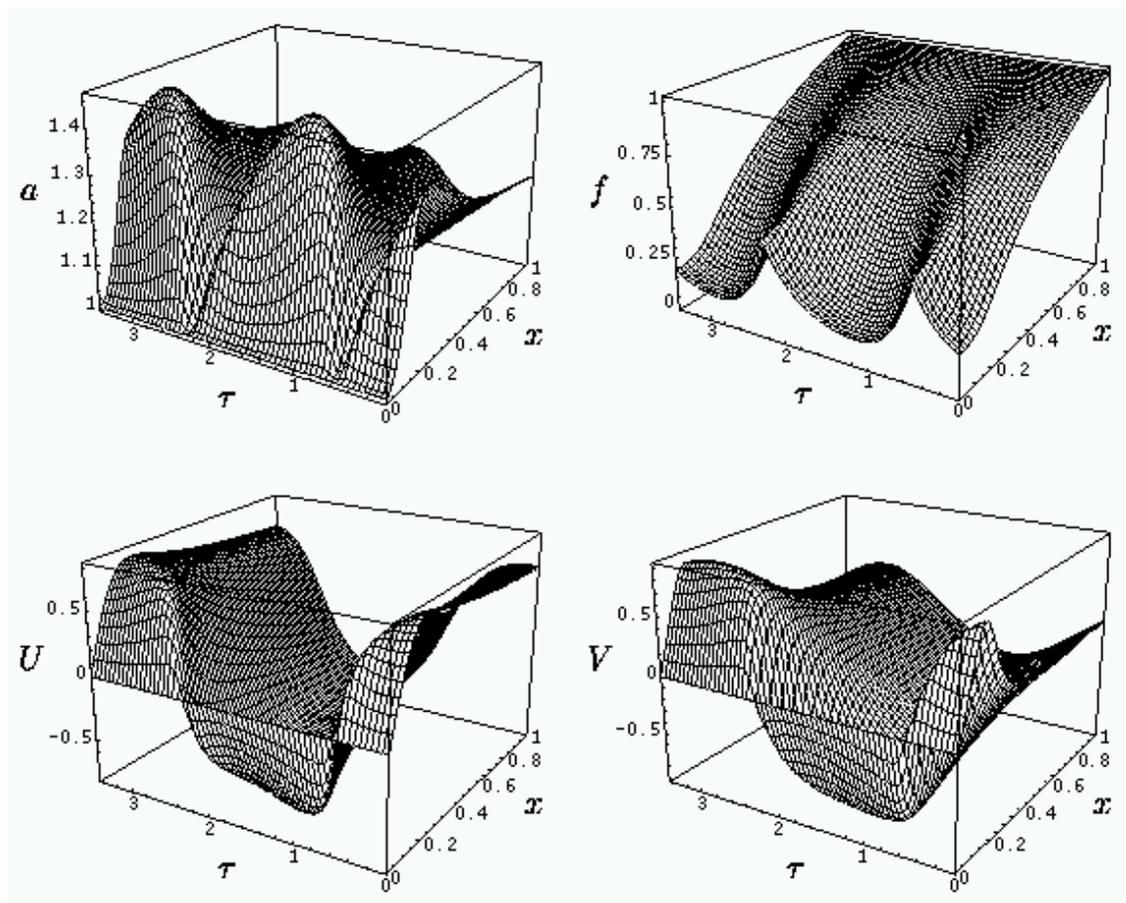}
\caption{\label{3dinnerplots}
Past patch fields on a single $\Delta$-period.
}
\end{figure*}

The natural choice for the coordinate of the center is $x_c=0$, but 
there is no preferred value for the past lightcone coordinate 
and we choose $x_p=1$. With the set of parameters
\begin{eqnarray}
N &=& 256, \nonumber \\
\xleft &=& 0.001, \nonumber \\
\xmid &=& 0.01 , \label {baseruninner} \\
\xright &=& 0.9999, \nonumber \\
\hmax &=& 7 \cdot 10^{-4}, \nonumber
\end{eqnarray}
our Newton's method converges to the free data plotted in 
Fig.~\ref{freedata} (see also table \ref{Fouriertable}), with a value
\begin{equation} \label{Deltavalue}
\Delta= 3.445452402(3),
\end{equation}
which improves the precision of our previous result $3.4453(5)$ 
more than four orders of magnitude. The metric and matter fields integrated
from those free data are given in Fig.~\ref{3dinnerplots}.

\begin{table*}
\caption{\label{Fouriertable} The first 8 nontrivial Fourier modes of 
the free data. (The error in the last digits is shown in brackets.)
Note that some of them have a relative precision better than $10^{-8}$.}
\begin{ruledtabular}
\begin{tabular}{c|llllll}
\quad $k$\quad $$ & Re $\hat f_{2k}(x_c)$ & Im $\hat f_{2k}(x_c)$ & Re $\hat\Psi_{2k+1}(x_c)$ & Im $\hat\Psi_{2k+1}(x_c)$ & Re $\hat U_{2k+1}(x_p)$ & Im $\hat U_{2k+1}(x_p)$ \\
\hline
0 & 0.2071909728(5) & 0 & 0.788624247(31) & -11.6194821(5) & 0.2962634507(4) & 0.0905094329(7) \\
1 & 0 {\rm \ \ by\ def.} & 0.0649057078(3) & 11.52753960(12) & 4.2699131(6) & -0.00901909339(14) & -0.02200156878(7) \\
2 & -0.02370998706(19) & -0.01438139603(19) & -9.12055613(23) & 7.5700655(6) & -0.002037853110(17) & 0.00159029448(4) \\
3 & 0.01347536638(18) & -0.00645699456(5) & -1.5735907(5) & -10.3634715(8) & 2.55305777(10) $10^{-4}$ & 1.73238389(4) $10^{-4}$ \\
4 & -0.001117368391(5) & 0.00883071824(10) & 8.2142645(16) & 3.3866919(5) & 1.12390710(11) $10^{-5}$ & -3.6797294(3) $10^{-5}$ \\
5 & -0.00432309030(6) & -0.00355712461(5) & -5.8935621(21) & 4.3873917(11) & -4.9301780(6) $10^{-6}$ & 8.2486(15) $10^{-9}$ \\
6 & 0.00345031858(11) & -0.00117027642(11) & -0.6075901(6) &  -5.9471050(31) & 1.884812(9) $10^{-7}$ & 6.14734(5) $10^{-7}$ \\
7 & -5.3677559(20) $10^{-4}$ & 0.00236648443(8) & 4.3606105(27) & 2.0264097(27) & 7.04459(5) $10^{-8}$ & -4.73630(8) $10^{-8}$
\end{tabular}
\end{ruledtabular}
\end{table*}

\begin{figure*}
\includegraphics[width=15cm]{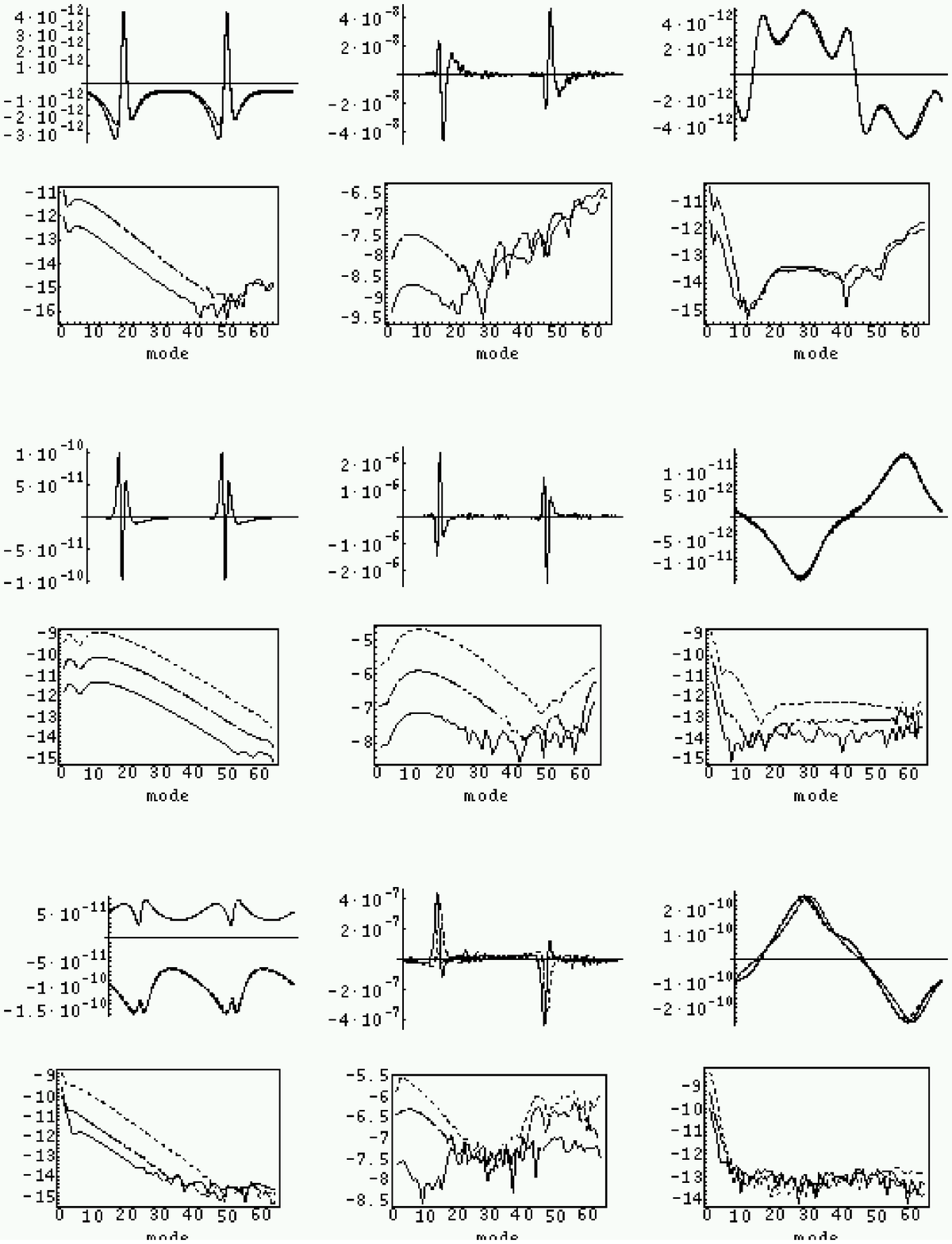}
\caption{ \label{convergence} Convergence figures for the past
patch. The three columns correspond to $f_c$, $\Psi_c$ and $U_p$,
respectively. The six rows are organized in three pairs, describing
convergence with respect to $\hmax$, $\xleft$ and $\xright$
respectively, when these three parameters are halved two or three
times. The first row of each pair shows consecutive differences of
fields, rescaled by factors 4, 16 and 5.2, respectively, so that they
coincide when converging with orders 2, 4 and 2.4, respectively. 
(Convergence with respect to $\xright$ is slower than the expected
order 3, in particular that of the very low frequency modes of $f_c$.)
The second row of each pair shows a $\log_{10}$ of the power spectrum
of those consecutive differences (without rescaling), to show the
different behaviour of the Fourier modes. Convergence of the high
frequency modes of $\Psi_c$ is worse than that of their low frequency
counterparts, as explained in the text.
}
\end{figure*}

\begin{figure}
\includegraphics[width=\columnwidth]{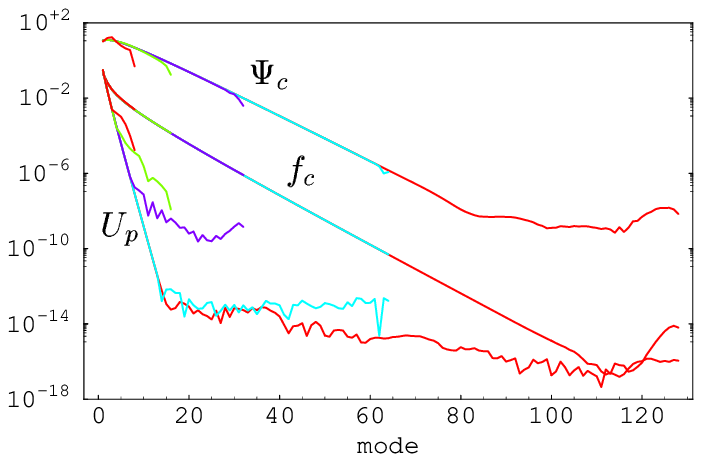}
\caption{\label{fourierfreedata}
Power spectra of the free data for 8, 16, 32, 64 and 128 modes.
Note the very different behaviors of the functions $f_c$ and $\Psi_c$
at the center from the function $U_p$ at the past lightcone.
}
\end{figure}

The error bars in (\ref{Deltavalue}) and table \ref{Fouriertable} are
estimated from the convergence properties of the code, which we now
explain in detail. In general, we have observed that we can reach
higher relative precision in the metric fields than in the matter
fields, because the former are essentially integrals of the latter.
Figs.~\ref{convergence} and \ref{fourierfreedata} contain all the
convergence information for the past patch, but in order to properly
discuss convergence issues, we first need to talk about an important
feature of the Choptuik spacetime.

\begin{figure*}
\includegraphics[width=15cm]{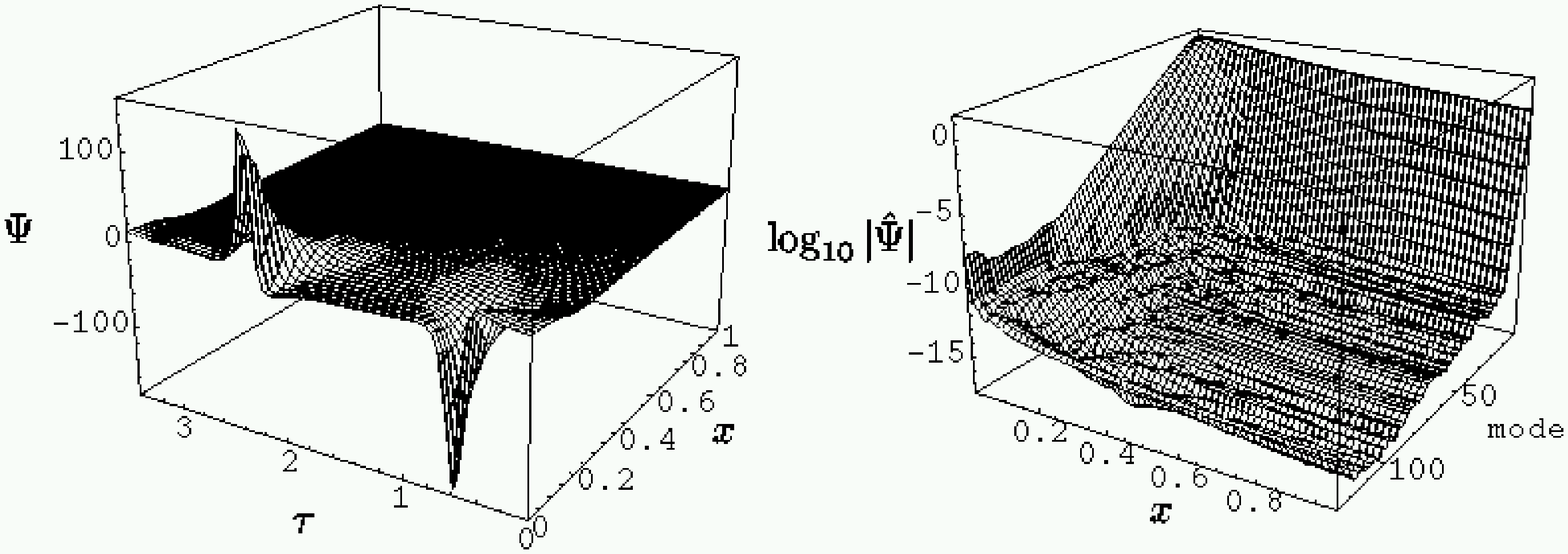}
\caption{ \label{psi3d}
$\Psi(\tau,x)$ and its power spectrum 
$\left|\hat\Psi_{2k+1}(x)\right|$. Note the difference in behaviour
between the regions below and above $x\sim 0.2$.
}
\end{figure*}

Fig.~\ref{psi3d} shows the field $\Psi(\tau,x)$ together with a
log$_{10}$ plot of the modulus of its Fourier transform in 
$\tau$. There is a clear difference in behaviour between the regions
above and below $x\approx 0.2$. (This difference is present in all our
fields, but it is particularly important in $\Psi$, as we will see.)
Near the center the function has large $\tau$-derivatives which require
many modes in the Fourier expansion to be resolved (see also 
Fig.~\ref{freedata}). Far from the center those derivatives are much
smaller and just a few modes are enough to achieve high resolution
results. This is reflected in a very slow decay of the Fourier transform
near $x\sim x_c$ and a fast decay for $x$ above $0.2$, although both are
exponential decays. It is important to emphasize that this exponential
decay is the best numerical evidence we have in support of the 
analytical character of the Choptuik spacetime, given the absence of a
mathematical proof of existence of an {\em analytical} solution to which
our numerical spacetime should be converging.

We have observed as well that the phases of the high frequency modes
$\hat Z_k$ tend to a linear dependence on $k$ at a given point $x$.
Therefore the {\it high frequency} behaviour of our fields $Z$ is
similar to that of the function
\begin{equation}
\sum_{k=-\infty}^\infty e^{-a|k|} \; e^\frac{4\pi i k \tau}{\Delta}
=\frac{\sinh a}{\cosh a - \cos\frac{4\pi\tau}{\Delta}},
\end{equation}
which has periodic sharp peaks of width $a\Delta/(4\pi)$ and height 
$2/a$ for small $a$. In the Choptuik spacetime the decay exponent $a$
ranges from $0.28$ for $f_c$ at the center to $2.04$ for $U_p$ at the
past lightcone. We have not found an explanation for this behavior,
but it seems to be of dynamical origin:
Arbitrary high-frequency perturbations of the correct free data at the
center decay towards larger $x$, and high-frequency perturbations at
the past lightcone grow when integrated towards the center,
probably due to $1/r$ factors in the equations of motion.

From a numerical point of view, this means that aliasing problems will
be important near the center $x=0$.  This forces us to use a large $N$
($N=512$ in Fig.~\ref{psi3d}). But then most of the modes
become essentially random for $x\gtrsim 0.2$ because they are well
below the error threshold given by roundoff error, estimated in
$10^{-14}$ in relative terms. This threshold generates the flat plateau
in the right panel of Fig.~\ref{psi3d}.
The main sources of roundoff error are threefold:
ODE integration along $x$, Fourier transforms, and inversion of a very
stiff matrix in Newton's method. Particularly important are the errors
in the modes of $U_p$ above $k=15$ because they propagate inwards and
get amplified, giving errors of relative order $10^{-9}$ in the matter
fields, mainly in $\Psi_c$. (See Fig.~\ref{psi3d} again.) 
This fact sets the limit of the maximum
accuracy that we can get in the results, with just double precision
numerics and using our code. We could use quadruple precision to
improve the precision in $U_p$ but the calculations would become
too slow. Alternatively, we could force the vanishing of those modes that we
believe must vanish, but we do not want to assume anything at all
about the result in advance.

We now check convergence with respect to the numerical parameters. As
one would expect, the final solution is completely insensitive to the
choice of intermediate fitting point $\xmid$, although the
convergence of Newton's method is faster when using smaller
values because the mismatches are larger. We choose $\xmid = 0.01$.

Several tests in simpler problems show that the ODE integrator in $x$
is perfectly fourth-order convergent. The first two rows of figure 
\ref{convergence} also show this fact, even though roundoff errors
slightly blur the point. Note that the modes of $U_p(\tau)$, after the
first 10, do not converge because they are already below our error
threshold (on the plateau in Fig.~\ref{psi3d}). Note also that
high-frequency modes of $\Psi_c$ do not converge for such small values
of $\hmax$. They do converge for larger values.

Convergence with $N$ is exponential as expected.
Fig.~\ref{fourierfreedata} shows the power spectra of the free data
for $N=32, 64, 128, 256, 512$. It clearly shows how much the results
are improved by doubling $N$ and how the plateau goes down each time,
until $N=256$, when errors in $U_p$ hit our error threshold.
The data for $N=512$ shows that we cannot improve the results any
further because we cannot decrease the errors in $U_p$. 
$f_c$ is then perfectly resolved down to machine precision,
but high-frequencies of the matter variables cannot be improved
near the center.
It is clear that using $N=256$ or $N=512$ we are not affected by
aliasing errors.

\begin{figure}
\includegraphics[width=\columnwidth]{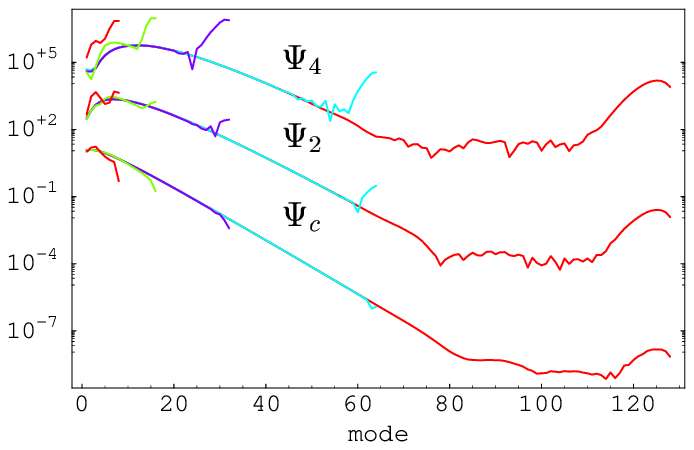}
\caption{ \label{psiders}
Power spectra of the functions $\Psi_c$, $\Psi_2$ and $\Psi_4$ for 8, 
16, 32, 64 and 128 modes. Aliasing problems show up as unphysically 
growing tails for the highest frequencies in each case. Working with 
only 32 modes, this introduces errors in $\Psi_2$ which are of the same
order of the amplitude of the most important modes. We cannot avoid
going to at least 64 modes (that is, $N=256$). $\Psi_4$ is presented for
illustration purposes only; we do not use it in the code.
}
\end{figure}

Finally, convergence with $\xleft$ and $\xright$ depends on the value
of $N$: As we said, we calculate the fields at $\xleft$ and $\xright$
using Taylor expansions:
\begin{eqnarray}
f(\tau,x)&=&f_c(\tau) + f_2(\tau)(x-x_c)^2
+ O\left[(x-x_c)^4\right] , \\
U(\tau,x)&=&U_p(\tau) + U_1(\tau)(x-x_p) \nonumber \\ 
&& + U_2(\tau)(x-x_p)^2 + O\left[(x-x_p)^3\right] ,
\end{eqnarray}
and so for the other fields. Therefore we expect fourth order
convergence with respect to $\xleft$ and third order with respect to
$\xright$. The coefficients of the Taylor expansions are obtained as
nonlinear combinations of the free data and their
$\tau$-derivatives. The latter are calculated multiplying the Fourier
modes $\hat{Z}_k$ by $ik$, which amplifies the high frequency
modes. Fig.~\ref{psiders} shows the Fourier transforms of
$\Psi_c,\Psi_2,\Psi_4$ obtained for different values of $N$. For
low $N$ our estimations of these coefficients are bad and we do not
see the expected convergence order in the expansions with $\xleft$ and
$\xright$, both because we are cutting off too soon in frequency,
leaving out modes which are important (see for instance the case of
$\Psi_2$ with $N=64$), and because of aliasing errors, which gives us
wrong estimations of the modes that we are including (see the
unphysical tails at the end of the functions).  The same phenomenon
happens on the past lightcone, but its effect is not so important.
That is the main reason why we need at least $N=256$
to get good results.  However, for higher $N$ we do see clear
convergence with the expected orders (with the exception of the very
low $k$ modes, whose convergence with respect to $\xright$ is slower
due to accumulation of high-frequency errors in $U_p$).
This is shown in Fig.~\ref{convergence}.  

We conclude that the code converges in the expected manner with
respect to all numerical parameters. Therefore we can estimate the
error of the base run (\ref{baseruninner}) with respect to any of the
parameters of the code as the difference between the base run and
another run with a refined value of that parameter. Those are
precisely the continuous lines in Fig.~\ref{convergence}. Truncation
errors from space and time discretization could be reduced to machine
precision. However, errors from the expansions at the singular points
cannot be brought down with our code below a limit which we
estimate (assuming no systematic error) between $10^{-8}$ and
$10^{-9}$ in relative terms. 
Therefore there is no point in reducing the other sources of
error below that limit, and the choice of parameters for our base run
(\ref{baseruninner}) reflects this.


\subsection{Outer patch}


In the entire outer patch the fields are much smoother in $\tau$ (as
they are already on the past patch for $x\gtrsim 0.2$). Therefore we
only need a few Fourier modes to get good precision. In fact $N=64$ is
enough to reach the maximum accuracy given by propagation of the
errors in the past patch. There is no reason to increase $N$ further.

\begin{table*}
\caption{\label{outerFouriertable} The first 8 nontrivial Fourier modes
of the outer patch free data. (The error in the last digit is shown in
brackets.) Note the very different relative precisions achieved in $\xi$
and $\hat U_\epsilon$. The former is insensitive to changes in the
parameters of the code, but not the latter.}
\begin{ruledtabular}
\begin{tabular}{c|llllll}
\quad $k$\quad $$ & Re $\hat \xi_{2k}$           &  Im $\hat \xi_{2k}$          & Re $\hat V_{2k+1}(x_f)$     & Im $\hat V_{2k+1}(x_f)$    & Re $\hat{\hat U}_\epsilon{}_{2k+1}$ & Im $\hat{\hat U}_\epsilon{}_{2k+1}$ \\
\hline
0                 &  1.322045988(6)                &  0                           &  -5.177664(12)$\cdot 10^{-4}$  &  3.157186(5)$\cdot 10^{-4}$  & -0.03844(5)                         & -0.250325(8)                        \\
1                 & -0.00492853319(12)             & -0.00430580184(14)               &  -1.5(3)$\cdot 10^{-10}$ &  -2.0(7) $\cdot 10^{-10}$  &  0.003883(8)                        & -0.0121478(20)                        \\
2                 & -1.1237092(5) $\cdot 10^{-5}$  &  2.89549492(5)$\cdot 10^{-4}$  &   -1.(4)     $\cdot 10^{-12}$ &  2.(15)     $\cdot 10^{-13}$ &  2.895(8)$\cdot 10^{-4}$            & -5.853(3)$\cdot 10^{-4}$            \\
3                 &  1.99072448(4)$\cdot 10^{-5}$  & -6.20831744(26) $\cdot 10^{-6}$  &   -1.(5)     $\cdot 10^{-13}$ & 0.(31)     $\cdot 10^{-14}$ &  1.748(5) $\cdot 10^{-5}$            & -2.5842(28) $\cdot 10^{-5}$            \\
4                 & -1.078319002(12)$\cdot 10^{-6}$ & -1.46908361(3) $\cdot 10^{-6}$  &  1.(30)     $\cdot 10^{-14}$ & -1.(6)     $\cdot 10^{-13}$ &  8.957(26)  $\cdot 10^{-7}$            & -1.0219(20) $\cdot 10^{-6}$            \\
5                 & -1.032893598(26) $\cdot 10^{-7}$  &  1.436706184(10) $\cdot 10^{-7}$  &  -0.(21)     $\cdot 10^{-14}$ & 0.(4)     $\cdot 10^{-13}$ &  4.126(10) $\cdot 10^{-8}$            & -3.920(10)  $\cdot 10^{-8}$            \\
6                 &  1.74888526(8) $\cdot 10^{-8}$  &  5.7084263(10) $\cdot 10^{-9}$  &  1.(25)     $\cdot 10^{-14}$ & 0.(4)     $\cdot 10^{-13}$ &  1.826(4) $\cdot 10^{-9}$            & -1.454(4) $\cdot 10^{-9}$            \\
7                 &  2.39356(8)  $\cdot 10^{-11}$ & -2.0053466(8) $\cdot 10^{-9}$  &  0.(3)     $\cdot 10^{-13}$ &  0.(3)     $\cdot 10^{-14}$ &  7.79(12)  $\cdot 10^{-11}$           & -5.21(26)  $\cdot 10^{-11}$
\end{tabular}
\end{ruledtabular}
\end{table*}

\begin{figure}
\includegraphics[width=\columnwidth]{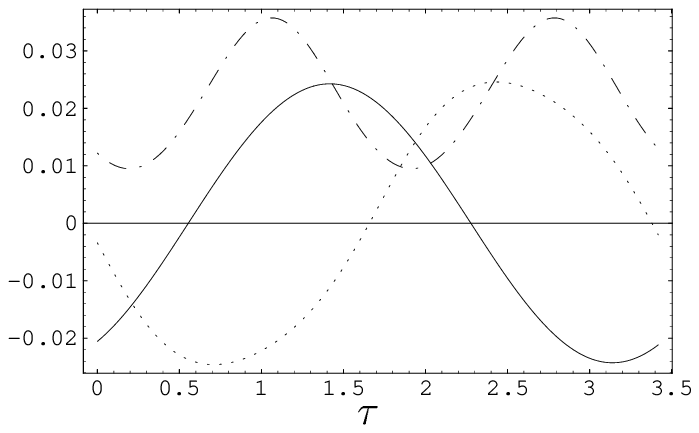}
\caption{ \label{freedata_outer}
Free data in the outer patch:  20$V_f$ (continuous line), 
$\hat U_\epsilon/20$ (dotted line) and $\xi-1.3$ (dash-dotted line).
Note that they are all quite smooth.
}
\end{figure}

With $x_p=-1$ and $x_f=1$ we choose these parameters for the numerical
evolution:
\begin{eqnarray}
N &=& 64, \nonumber \\
\xleft &=& -0.9999, \nonumber \\
\xmid &=& -0.9, \label {baserunouter} \\
\xright &=& 0.9999, \nonumber \\
\hmax &=& 0.001, \nonumber
\end{eqnarray}
Now the free data are the metric function $\xi(\tau)$ and the matter
functions $V_f(\tau)\equiv V(\tau,x_f)$ and $\hat U_\epsilon(\tau)$ at
the future lightcone. (Note that the function $V_f$ was called $V_0$ in
Subsection \ref{section:singularCH}.)The results are given in 
table \ref{outerFouriertable} and Fig.~\ref{freedata_outer}, with a 
final value
\begin{equation}
\label{epsilon_value}
\epsilon = 1.4710439(8)\cdot 10^{-6} ,
\end{equation}
clearly different from zero. We firmly believe that this is neither a
numerical artifact nor a consequence of our expansion around the
future lightcone. After showing convergence of the code in the outer
patch, we dedicate most of this subsection to supporting this claim.

\begin{figure*}
\includegraphics[width=15cm]{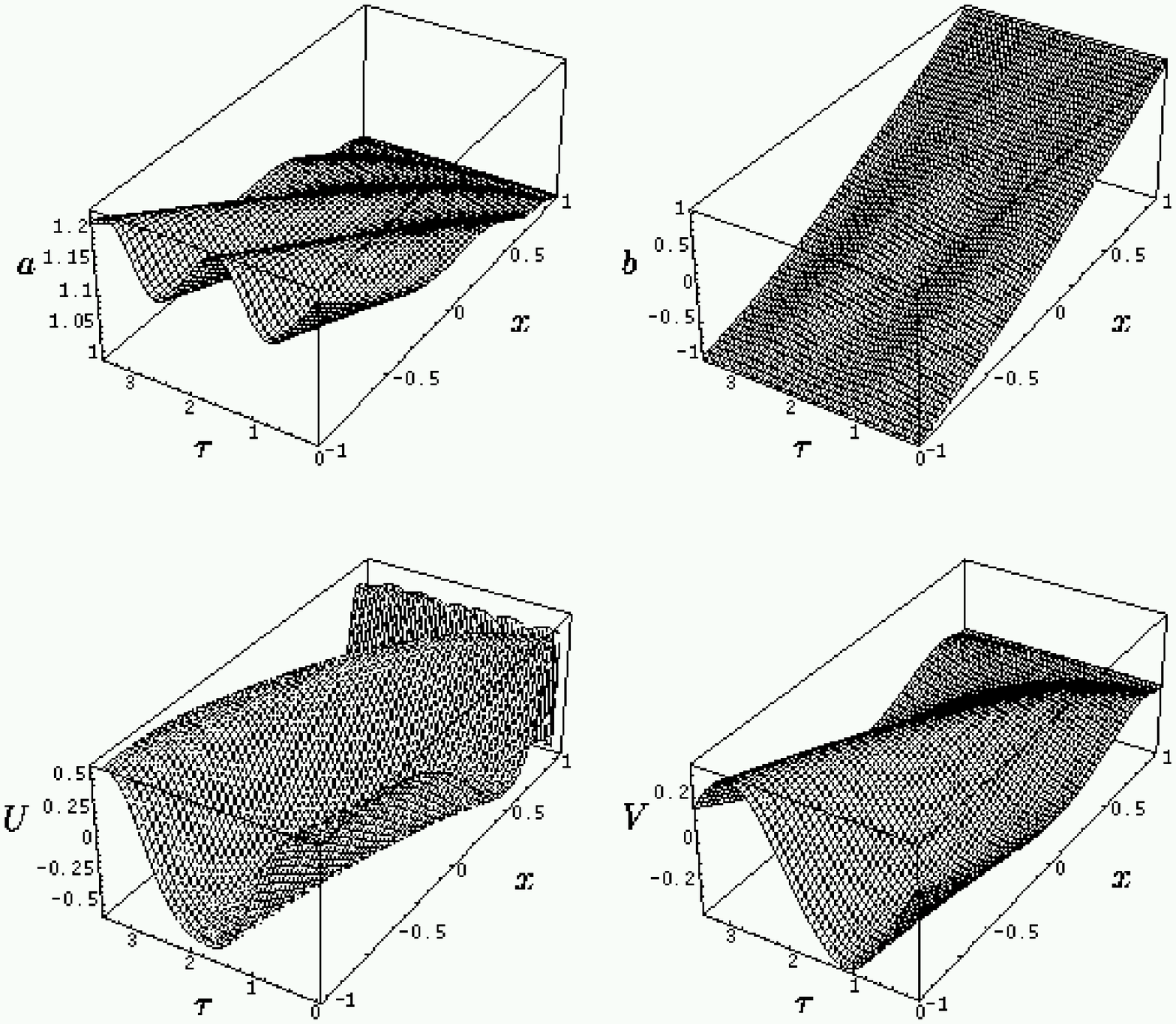}
\caption{ \label{choptuon_outer}
Outer patch fields on a single $\Delta$-period.
}
\end{figure*}

\begin{figure}
\includegraphics[width=\columnwidth]{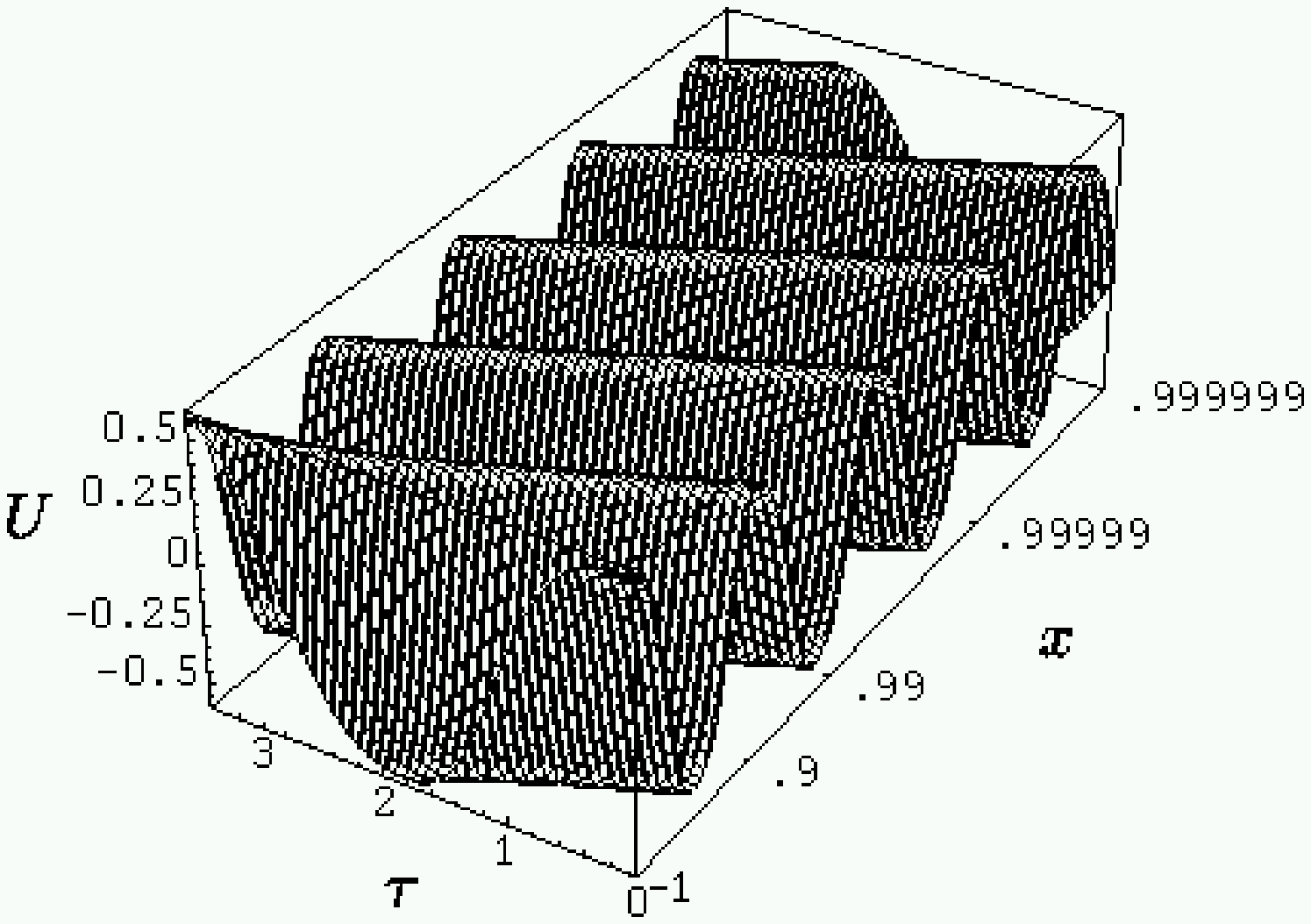}
\caption{ \label{Ulog} $U$ in the outer patch using a logarithmic
$x$-axis. The oscillations are clear, but not their slow decay.  }
\end{figure}

The integrated functions are shown in Fig.~\ref{choptuon_outer}. The
runaway of characteristics is not apparent in this figure, because all
but the first oscillations are piled up in the region between $x=0.95$
and $x=1$. Fig.~\ref{Ulog} shows $U$ using a logarithmic $x$-axis.

\begin{figure*}
\includegraphics[width=15cm]{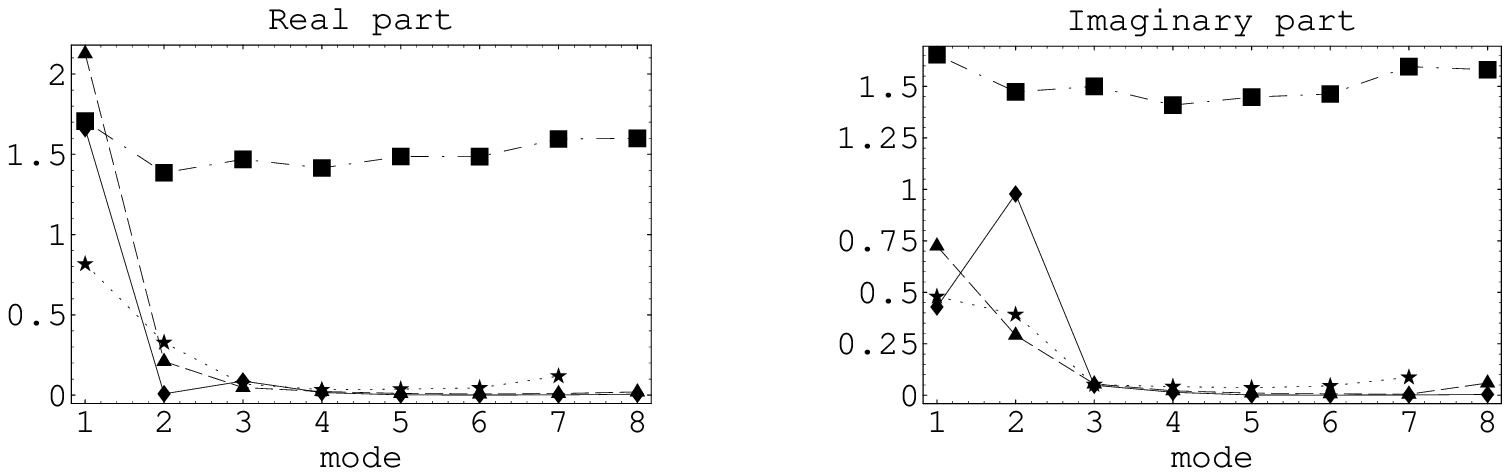}
\caption{ \label{in2outerror}
Variation of the results of the outer patch under changes of the input
data from the past patch. On the left we represent 
$||\delta f||_\infty/\delta p_k$ where $\delta f$ is the change in 
$10^5 \epsilon$ (continuous line) or in one of the functions $10^2 V_f$
(dotted line), $\xi$ (dashed line) or $\hat U_\epsilon$ (dot-dashed 
line), under a change $\delta p_k$ in the real part of the mode $k$ of 
$U_p$.  On the right we show the same for the imaginary part of the
modes of $U_p$. We lack enough precision to calculate the last point
of the $V_f$ curves.
}
\end{figure*}

We first analyze the issue of error propagation from the past patch to
the outer patch. We need to find out how small variations in $\Delta$
and $U_p$ change the results of the outer patch shooting. Assuming that
the former are small, we calculate first derivatives of the latter.
The variations of the outer patch free data with respect to $\Delta$
are:
\begin{eqnarray}
\frac{\delta \epsilon}{\delta\Delta} = 5.4\cdot 10^{-6} , &\qquad&
\frac{||\delta V_f||_\infty}{\delta\Delta} = 0.0020 , \\
\frac{||\delta \hat U_\epsilon||_\infty}{\delta\Delta} = 0.15 , &\qquad&
\frac{||\delta \xi||_\infty}{\delta\Delta} = 0.028 .
\end{eqnarray}
On the other hand, Fig.~\ref{in2outerror} shows the maximum 
variations of the free data with respect to changes of the Fourier
modes of $U_p$. We see that $V_f$ and $\xi$ are only sensitive to the
very low frequency modes of $U_p$, but $\hat U_\epsilon$ changes with
every mode. In any case, every derivative is small enough:
The largest error bars come from the uncertainty in $\Delta$, and
then from those of the first two modes in $U_p$. The rest of the
modes are practically irrelevant for error analysis. This sets the
maximum accuracy that we can achieve on our final results. Assuming
quadratic error propagation 
it is
\begin{eqnarray}
\delta_{\rm max}\epsilon &=& 5\cdot 10^{-13} , \\
||\delta_{\rm max} V_f||_\infty &=& 2\cdot 10^{-9} , \\
||\delta_{\rm max} \hat U_\epsilon||_\infty &=& 4\cdot 10^{-8} , \\
||\delta_{\rm max} \xi||_\infty &=& 2\cdot 10^{-8} .
\end{eqnarray}
It is still very good due to the tendency of small perturbations to
decay when integrating towards larger $x$, and, as we said, achievable
already with $N=64$.

\begin{figure*}
\includegraphics[width=15cm]{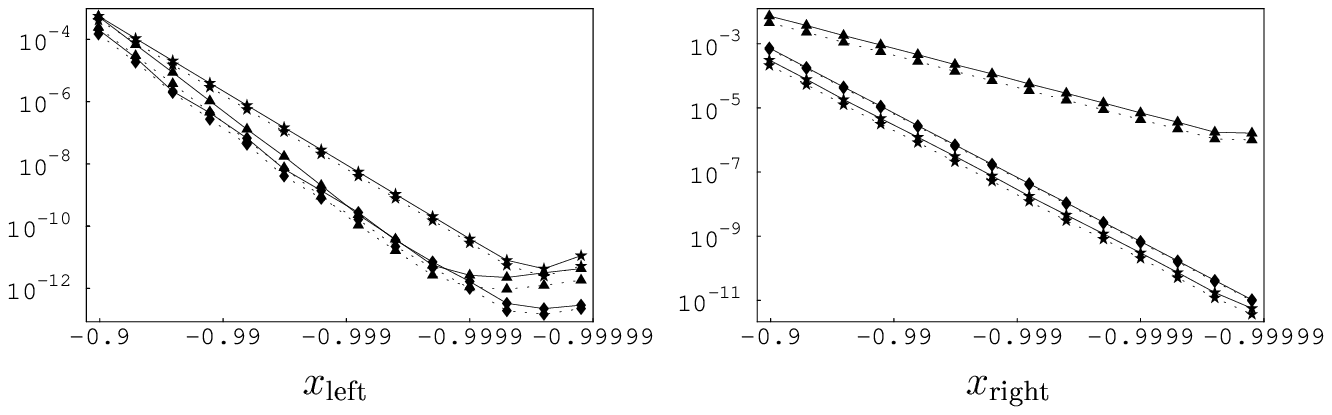}
\caption{ \label{outerconv}
Convergence of the free data on the outer patch. These plots display
norms of differences between consecutive results: $\infty$-norm
(continuous line) and 2-norm (dotted line). On the left, convergence
with respect to $\xleft$ shows order 3.0 for $\hat U_\epsilon$
(triangles) and $\xi$ (diamonds) and order 2.4 for $V_f$ (stars). On the
right, convergence with respect to $\xright$ shows order 2.0 for
$\xi$ and $V_f$ and order 1.0 for $\hat U_\epsilon$.
}
\end{figure*}

We now analyze convergence in the outer patch. 
Convergence with $\hmax$ in the IRK2 method shows perfect fourth order
again. Convergence with $\xleft$ is approximately third order as
expected because we expand around the past lightcone with a
second-order Taylor series. See Fig.~\ref{outerconv}.
Finally, we have performed calculations expanding around $x_f$ using
only the order zero terms and including the first order terms.

$V_f$ and $\xi$ converge with $\xright$ to first order when the
expansion around the CH is truncated at $O(|y|^\epsilon)$, and converge
to second order when the expansion is truncated at
$O(|y|^{1+3\epsilon})$. This is the expected behavior. However, at the
same time $\hat U_\epsilon$ converges to first order in both cases
(see Fig.~\ref{outerconv}).
This indicates that adding the terms of order $O(|y|^{1+k\epsilon})$
(with $k=0,1,2,3$) to the expansion still leaves some 
$O(|y|^{1+k\epsilon})$ error. We are confident that this is not a simple
algebraic mistake in the expansion. We note that the excess error in the
periodic function $\hat U_\epsilon$ is entirely an error in its overall
phase. We therefore suspect intuitively that the runaway phase
$U\sim \hat U_\epsilon(\tau-\ln|y|)$ of the solution is to blame, but we
have not been able to formulate this idea consistently. 

\begin{figure*}
\includegraphics[width=15cm]{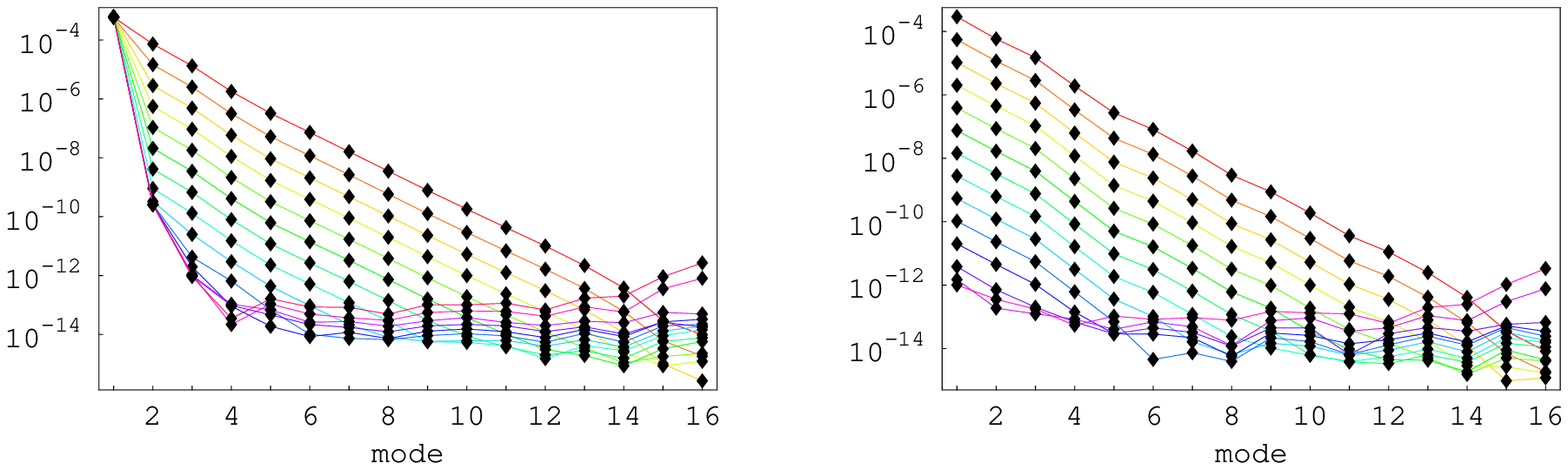}
\caption{ \label{Vfconv}
Convergence of $V_f$. Left figure: $\log_{10}|\hat V_f{}_{2k-1}|$ for 15
values $\xleft=$-0.8976, -0.9488, -0.9744, -0.9872, -0.9936, -0.9968,
-0.9984, -0.9992, -0.9996, -0.9998, -0.9999, -0.99995, -0.999975,
-0.9999875, corresponding from top to bottom, respectively.
It seems that all but the first two modes converge to amplitudes below
our error threshold when $\xleft$ approaches -1, even though
high-frequencies become unstable for the last values of $\xleft$.
Right figure: modulus of the
Fourier transform of consecutive differences in the left figure.
Convergence is very clear in every mode (including the first one).
}
\end{figure*}

In order to show that $\epsilon$ is really different from zero, we
have to analyze the behavior of the function $V$ with respect to
$\xleft$ and $\xright$. The function $V_f$ has a very rapidly decaying
Fourier spectrum, as shown in Fig.~\ref{Vfconv}. As $x_{\rm left}\to
x_p$, all its Fourier modes converge to zero except for the first two,
and the amplitude of the second mode is more than six orders of
magnitud below that of the first one (see also 
Table~\ref{outerFouriertable}).

\begin{figure}
\includegraphics[width=\columnwidth]{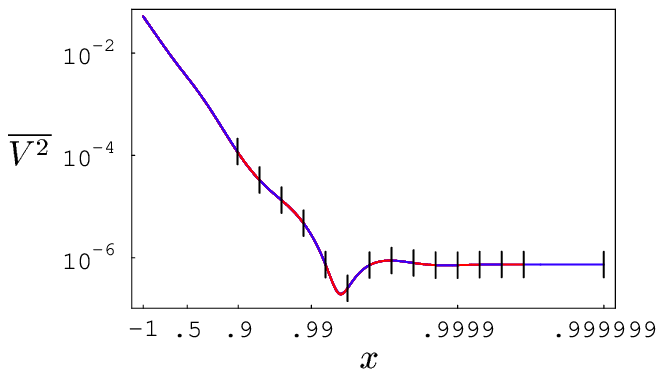}
\caption{ \label{V2conv}
Average of $V^2$ on $\tau$ for different evolutions from the same
$\xleft=-0.9999$ to 15 different values of $\xright$ (those of table
\ref{eps_conv}). The agreement between the solutions with small
$\xright$ and those with $\xright$ very close to $x_f=1$ implies that
the final almost constant value is neither a numerical artifact nor a
consequence of our ansatz.
}
\end{figure}

Fig.~\ref{V2conv} shows $\overline{V^2}(x)$ for several
evolutions from the same $\xleft=-0.9999$ to 15 different values of
$\xright$. The agreement is very good. This shows that our expansion
around the future lightcone (including the singular terms given in
the appendix) captures the behavior of the solution.

We conclude that, with errors of order $10^{-9}$,
\begin{equation}
\label{V_f_numerical}
V_f(\tau)=-1.2128648(22)\cdot 10^{-3} \cdot 
\cos\left(\frac{2\pi\tau}{\Delta} +0.5475726(13)\right) .
\end{equation}

\begin{table}
\caption{\label{eps_conv}
Convergence of $10^6\epsilon$ with $\xright$ using only the lowest order
terms in the regular expansion plus only the $O(y^\epsilon)$ in U
(first column) and using the full next order as well (second column).
Convergence is much faster in the second case, as expected, but it is
clear that both converge to the same number, within our error bars.
Recall that digits starting from the fifth decimal are not relevant
due to propagation of errors from the past patch. They are shown
in the second column to make convergence clear.
}
\begin{ruledtabular}
\begin{tabular}{c|cc}
$\xright$ & zeroth order & first order \\
\hline
0.8976     & 238.46 & 0.87085     \\
0.9488     & 66.166 & 1.36313     \\
0.9744     & 26.743 & 1.49818     \\
0.9872     & 9.4840  & 1.48105     \\
0.9936     & 1.4699  & 1.47058     \\
0.9968     & 0.5058  & 1.470305    \\
0.9984     & 1.4386  & 1.470949    \\
0.9992     & 1.7639  & 1.4710727   \\
0.9996     & 1.5837 & 1.4710534   \\
0.9998     & 1.4355 & 1.47104333  \\
0.9999     & 1.4332 & 1.47104318  \\
0.99995    & 1.4687 & 1.471043825 \\
0.999975   & 1.4799 & 1.471043941 \\
0.9999875  & 1.4743 & 1.471043921 \\
0.999999   & 1.4713 & 1.471043911 
\end{tabular}
\end{ruledtabular}
\end{table}

Finally, it is interesting to see that going to $O(|y|^{1+3\epsilon})$
is not essential for obtaining an accurate result. Table
\ref{eps_conv} compares the results for $\epsilon$ using a
zeroth-order expansion (that is, we only include the terms of orders
$y^0$ and $|y|^\epsilon$) with those from the first order expansion
(which also includes orders $y$ and $|y|^{1+k\epsilon}$). It is clear
that they both converge to the same number, even though with very
different rates of convergence.


\section{Continuation across the Cauchy horizon: the future patch}
\label{section:continuation}



\subsection{The continuation with a regular center}


We cannot continue the solution as flat empty Minkowski spacetime
after the Cauchy horizon because we have a small amount of ingoing
radiation there, given by $V_f(\tau)$. The simplest continuation to
look for is one with a regular timelike center, so that the conformal
diagram is the same as for Minkowski spacetime. In this case both $U$
and $V$ are small on the entire future patch, and we can obtain an
approximate solution in perturbation theory around Minkowski space, using the
magnitude $\epsilon^{1/2}$ of $V_f$ as the small parameter. To
leading order in $\epsilon$ we obtain the d'Alembert solution on flat
spacetime:
\begin{eqnarray}
f &=& 1+O(\epsilon) , \\
a &=& 1+O(\epsilon) , \\
U &=& \epsilon^{1/2} \left[-F'(\hat\tau) -
\frac{F(\hat\tau)+G(\tau)}{x}\right] +O(\epsilon^{3/2}) , 
\label{Upert} \\
V &=& \epsilon^{1/2} \left[-G'(\tau) +
(1+x)\frac{F(\hat\tau)+G(\tau)}{x} \right] +O(\epsilon^{3/2}),
\label{Vpert}
\end{eqnarray}
with $\hat\tau\equiv\tau-\ln(1+x)$. In order to match the null data on
the Cauchy horizon $x=-1$, we need $-\epsilon^{1/2}
G'(\tau)=V_f(\tau)$. Recall that $V_f(\tau)$ is given numerically by
(\ref{V_f_numerical}). If we want to have a regular solution at the
center we need $F(\tau)=-G(\tau)$. The solution is then completely
determined, and it is clear that a nonlinear solution exists of
which this is the leading order, and which can be found numerically.

\begin{figure*}
\includegraphics[width=15cm]{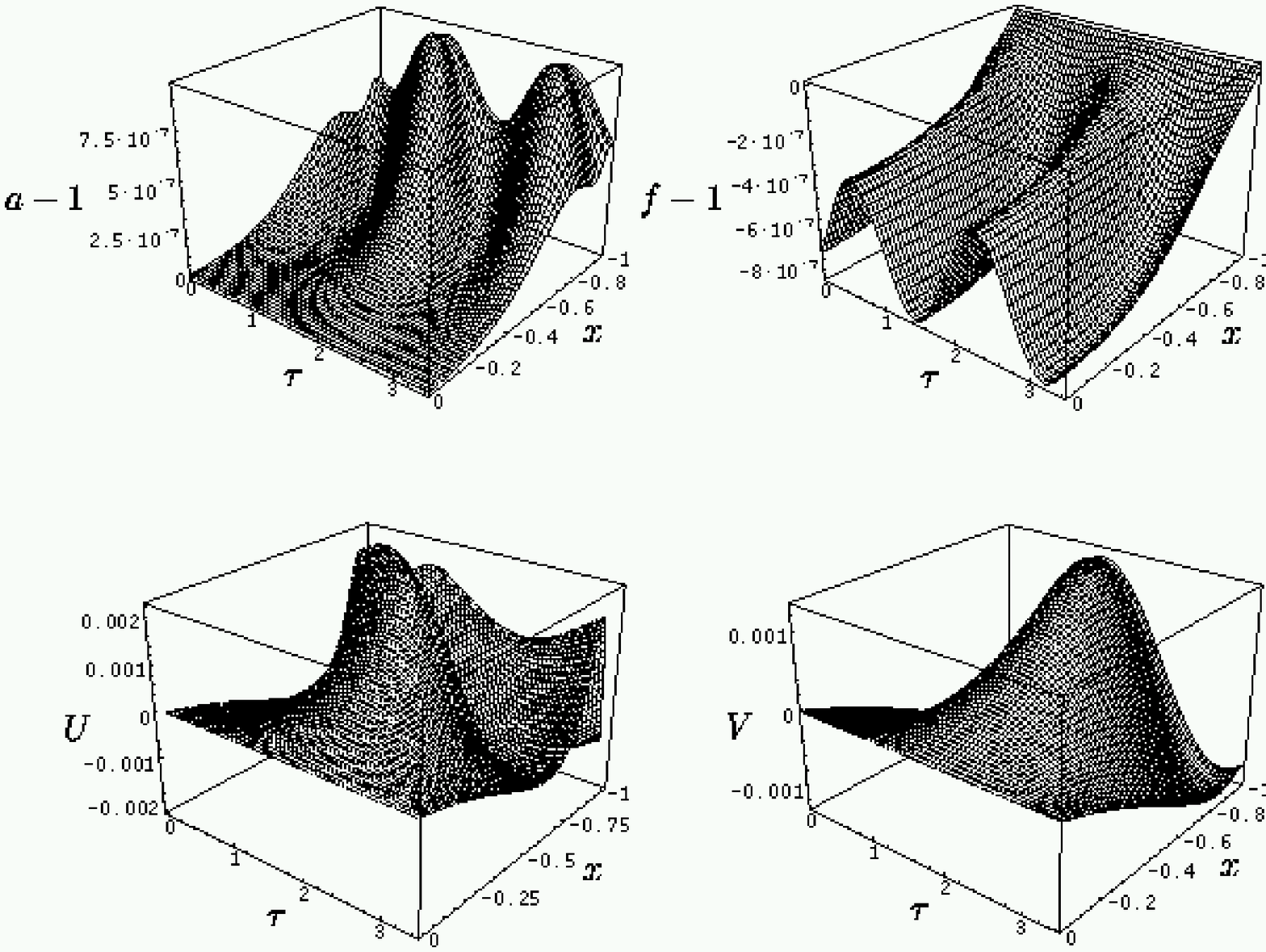}
\caption{\label{choptuon_future} Future patch fields on a single
$\Delta$-period for the (probably unique) continuation with a regular
timelike center. Note that $a$ and $f$ are so close to their flat
spacetime values that we plot their difference from 1, rather than the
fields themselves.  }
\end{figure*}

Because the null data $V_f$ on the Cauchy horizon are DSS, it appears
highly unlikely that there is another continuation with a regular
timelike center that is not DSS. We have obtained the (probably)
unique regular continuation by shooting from expansions around the
Cauchy horizon and a regular center. The free data for the shooting
algorithm (given by $\hat{U}_\epsilon$ at the CH and $f$, $U_{,x}$ at
the center) were obtained from the flat spacetime approximation. In
this case we use an IRK1 integrator and
\begin{eqnarray}
N &=& 16, \nonumber \\
\xleft &=& -0.999, \nonumber \\
\xmid &=& -0.2, \label {baserunfuture} \\
\xright &=& -0.001, \nonumber \\
\hmax &=& 2.5\cdot 10^{-4}, \nonumber
\end{eqnarray}
with very good convergence, that we do not show again. The fields $a$, 
$f$, $U$ and $V$ are shown in Fig.~\ref{choptuon_future}. 

\subsection{All other continuations}


We now consider the other possible continuations, in particular those
of Figs.~\ref{figure:timelikesingular} and
\ref{figure:spacelikesingular}. Before performing a numerical search
of the possibilities in the $\hat U_\epsilon$ space, we study the
equations of motion in the future patch. We proceed in four steps:

\paragraph{A necessary condition for another
SSH is $a^2>2$.} In order to obtain
Fig.~\ref{figure:spacelikesingular},
or any even more exotic
continuation, we must have a self-similarity horizon before the
central singularity occurs at $x=x_s=0$.

A self-similarity horizon is a DSS line $x=x_h(\tau)$ (periodic) where
$A=0$. The only factor in $A=-4a^2 f(f+x)$ that can vanish (while the
metric is regular) is $f+x$. We have
\begin{equation}
(f+x)_{,x}=\frac{(a^2-1)f}{x}+1
\end{equation}
and so
\begin{equation}
\left.(f+x)_{,x}\right|_{f+x=0}=2-a^2 .
\end{equation}
At the Cauchy horizon $f+x=0$ and $(f+x)_{,x}\simeq 1$ because $a^2-1$
is small, and so $(f+x)>0$ and $A<0$ at least to the immediate future of
the Cauchy horizon. If there are more self-similarity horizons to the
future of the Cauchy horizon, we must have $f+x=0$ again there, and
therefore $(f+x)_{,x}<0$ in some intermediate region. $a^2$ must
increase from $a^2\simeq 1$ to $a^2>2$ in order to achieve this.

\paragraph{Once $a^2<1$ for any $\tau$, $a^2\to 0$
for that $\tau$.} $a$ is given by the constraint (\ref{a_constraint})
which reduces at the Cauchy horizon to
\begin{equation}
(a^{-2})_{,\tau}=(1+2V_f^2)a^{-2}-1,
\end{equation}
and this means that $1 < a < 1+\epsilon$ there. Now $a$ obeys
the evolution equation
\begin{equation}
-2x(\ln a)_{,x}=a^2-1-2U^2.
\end{equation}
Recall that in the future patch $x<0$ and increases towards the
future. Therefore $a=1$ is a repeller in the absence of matter, but
any outgoing radiation $U$ drives $a$ to smaller values. Hence, once
$a$ becomes smaller than 1 for some value of $\tau$, it will
afterwards decrease to 0 for that $\tau$ whatever happens.
Because $\mu=2m/r=1-a^{-2}$, this means that we are in a negative mass
regime and remain there until we reach the central singularity $x=0$,
which is therefore timelike and has negative infinite mass at least
for this $\tau$.  The singularity cannot be reached with a finite
value of $0<a^2<1$.

\paragraph{Is $a^2>1$ at the singularity possible at least for some
values of $\tau$?} If we choose $\hat U_\epsilon$ much larger than
$O(\epsilon^{1/2})$, the $U^2$ term will drive $a$ to $a<1$ almost
immediately. However, we know that the regular solution does not have
negative mass and therefore it seems plausible that functions $\hat
U_\epsilon$ close to the regular case could generate a singularity
with positive mass. We shall now argue that this is not true.

$a$ can only increase from its CH value if $|\hat U_\epsilon|$ is very
small. In order to explore this regime we return to the approximation of
perturbing around Minkowski spacetime, 
but now dropping the assumption that $F(\tau)=-G(\tau)$.
The result can be summarized as:
\begin{eqnarray}
\label{pertsing}
a(\tau,x) &=& 1 + \epsilon \; x^2 a_{\rm reg}(\tau,x) \nonumber \\
&-& \epsilon \; x^{-2}\left[F(\tau)+G(\tau)\right]^2  \\
&+&  \epsilon \; O(x^{-1}) + O(\epsilon^{3/2}) . \nonumber
\end{eqnarray}
The function $a_{\rm reg}$ is positive and always smaller than 2, and
is independent of $F$. All singular terms vanish for $F=-G$. For every
other $F$, the function $a$ becomes smaller than 1, which
corresponds to negative mass, as the center $x=0$ is approached.
This is due to the divergent terms $(F+G)/x$ in (\ref{Upert}) and
(\ref{Vpert}). The crucial point is that $a^2$ is integrated from
$-U^2$ and therefore the term in the perturbation expansion that
makes the mass nonzero has coefficient $-(F+G)^2$, and so the mass
cannot be positive. 
The negative mass regime is reached while perturbation theory still
applies, and we have seen that afterwards the mass must decrease
indefinitely.

With these arguments we have ruled out the possibility of having
positive mass at the singularity for very large $\hat U_\epsilon$ and
for very small (in particular, close to the regular case) 
$\hat U_\epsilon$. However we could be missing some intermediate regime
where perturbation theory does not apply. For these intermediate cases
we observe that the minimum of $a$ is far from 1 (and therefore we
cannot apply perturbation theory), but its maximum is positive and
close to 1, and only becomes negative in the vicinity of the
singularity. There could be solutions where $a$ is positive at the
singularity for some values of $\tau$.

\begin{figure}
\includegraphics[width=\columnwidth]{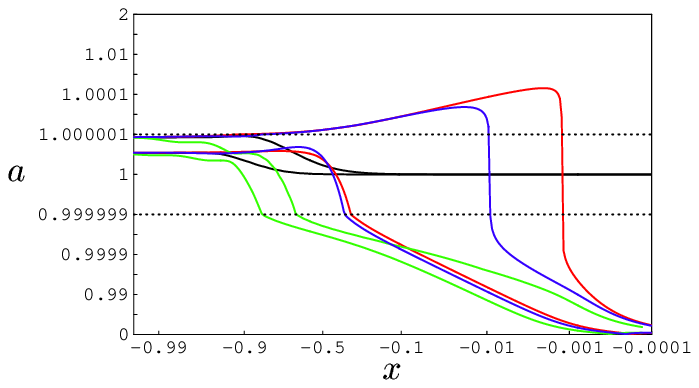}
\caption{ \label{negativemass}
Maximum and minimum of $a$ for different initial data functions 
$\hat U_\epsilon$ from numerical evolutions of the nonlinear equations.
Note that the $x$ axis becomes logarithmic when approaching the
boundaries, while the $a$ axis is linear between the dotted lines at 
$a=1-10^{-6}$, $a=1+10^{-6}$, and logarithmic elsewhere.
The initial $\hat U_\epsilon$ have been obtained multiplying the regular
data (black) by different constants: 0 (red), -1 (blue) and 5 (green).
Clearly, the larger $\hat U_\epsilon$ is, the sooner $a$ decays.
}
\end{figure}

Numerically, however, we do not find that. We have performed a large
number of numerical evolutions in this intermediate regime starting
from $\hat U_\epsilon= 2^n \,\hat U_\epsilon{}_{\rm
reg}(\tau+m\Delta/8)$, for $n=-5,...,10$ and $m=0,...,7$. Even though
it is difficult to evolve the system near the singularity, we always
observe a final decay of $a$ to 0. See
Fig.~\ref{negativemass}. Therefore we believe that, starting from
small null data $V_f$ at the CH it is not possible to form another
SSH. Therefore, either we have a regular center or a timelike 
singularity, and this singularity always has negative mass.
In the approach to the center with $a^2\to
0$ we always have $U+V\to 0$ with both $U,V$ finite. It should be
stressed that this scenario is a consequence of the small amplitude of
the null data $V_f$ on the Cauchy horizon in the Choptuik solution. We
have performed other evolutions starting from large null data on the
Cauchy horizon
(not $V_f$ of the Choptuik solution) where $a^2$ increases beyond 2.

\paragraph{How is the singularity approached?}

We find that in all numerical continuations $a\to 0$, $f\to \infty$,
$U$ and $V$ tend to finite values and $U\to -V$ as the singularity
$x=0$ is approached. If we assume that $a\to 0$ and $U\to U_0(\tau)$
as $x\to 0$ then the $a_{,x}$ and $f_{,x}$ equations to leading order
give
\begin{eqnarray}
f&\simeq& f_0(\tau)\, x^{-1}, \\
a^2&\simeq& a_0^2(\tau) \, |x|^{2\alpha(\tau)}, 
\qquad \alpha(\tau)\equiv
\frac{1}{2}+U_0(\tau)^2.
\end{eqnarray}
Making the ansatz that
\begin{eqnarray}
U&\simeq&U_0(\tau)+U_2(\tau)\, x^2 \nonumber  \\ 
&& + \, U_{2l}(\tau)\, x^2\ln |x|
+ U_{\alpha}(\tau) \, |x|^{2\alpha(\tau)} , \\
V&\simeq& V_0(\tau)+V_2(\tau)\, x^2 \nonumber  \\
&& + \, V_{2l}(\tau)\, x^2\ln |x|
+ V_{\alpha}(\tau) \, |x|^{2\alpha(\tau)} ,
\end{eqnarray}
we find from the $U_{,x}$ and $V_{,x}$ equations to the leading
order, $O(\ln |x|)$ that
\begin{eqnarray}
V_0&=&-U_0, \\
V_2&=&U_2+{1\over 2}f_0^{-1}U_0', \\
U_{2l}=V_{2l}&=&-\frac{1}{2} f_0^{-1}U_0', \\
U_\alpha=-V_\alpha&=&-\frac{1}{2\alpha}a_0^2U_0.
\end{eqnarray}
(The expansion also holds in the special case $\alpha=1$.)
The next
order, $O(1)$, gives
\begin{equation}
\label{constraint}
\ln(a_0^2)'+1+2U_0^2 +8f_0U_0U_2+2U_0U_0'=0.
\end{equation}
The metric in the future patch contains a residual gauge freedom worth
one periodic function of $\tau$. Near the singularity $x=0$, we can
fix this gauge freedom by setting $f_0(\tau)$ to an arbitrary
value. (In our numerical evolutions starting at the Cauchy horizon,
the gauge has of course been fixed already by setting $f=1$ 
at the Cauchy horizon.)
$U_0(\tau)$ and $U_2(\tau)$ are then physical free data which
determine $a_0(\tau)$ and all other coefficients of the
expansion. (Eq. (\ref{constraint}) can be solved for $a_0$ only if the
right-hand side has vanishing average, which means that $f_0$ cannot
be set completely freely.) This expansion is
therefore generic in the sense of depending on two free functions
after the gauge has been fixed.

\begin{figure*}
\includegraphics[width=15cm]{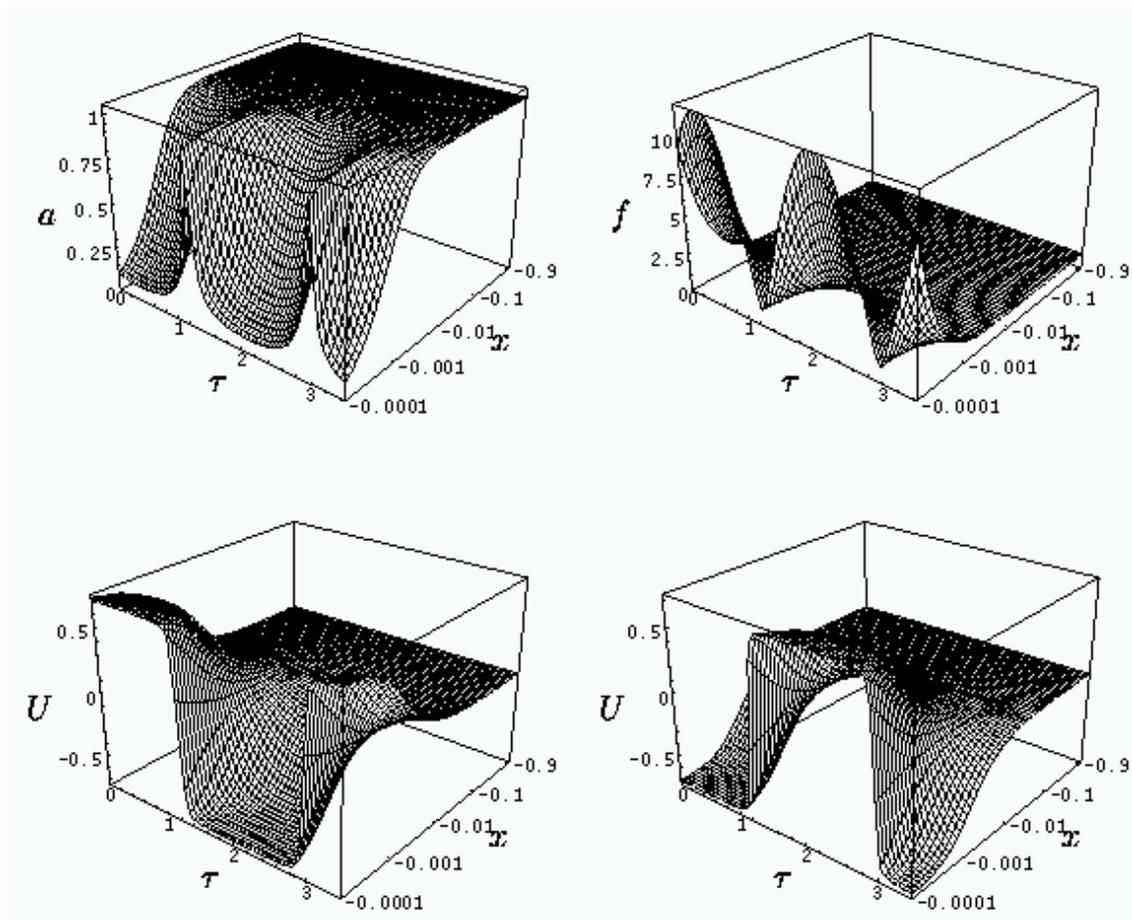}
\caption{ \label{choptuon_singular}
Future patch fields on a single $\Delta$-period for a singular
continuation of the Choptuik spacetime.
}
\end{figure*}

The behavior just described is what we observe numerically
for large values of
the free data $\hat U_\epsilon(\tau)$ at the Cauchy horizon. For small
values of $\hat U_\epsilon(\tau)$ we find $U_0(\tau)\simeq \pm
1/\sqrt{2}$, that is $\alpha(\tau)\simeq 1$. Note that by our ansatz
of exact DSS with $\kappa=0$, $U$ must be an odd function of $\tau$
with zero average. We observe in fact that $U(\tau,x)$ goes to a
fundamental frequency square wave of amplitude $1/\sqrt{2}$, that is,
in half of each $\tau$-period $U\to 1/\sqrt{2}$, and $U\to
-1/\sqrt{2}$ in the other half. This is shown in
Fig.~\ref{choptuon_singular}. 

As the center $x=0$ is approached with $U_0=\pm 1/\sqrt{2}$, we
observe empirically that the $\tau$-derivatives become dynamically
negligible. This means that different values of $\tau$ effectively
decouple, that at each point $\tau$ the evolution equations become an
ODE system in $x$, while the constraint becomes algebraic, and that
the spacetime becomes locally CSS. It also means that the evolution
equations become ``velocity-dominated'' in the sense that all
derivatives transversal to the singularity (here in spherical
symmetry, this is only the $\tau$-derivative) become dynamically irrelevant
compared to the one derivative running into the singularity (here, the
$x$-derivative). It is known that generic spacelike singularities in
general relativity with massless scalar field matter are
velocity-dominated \cite{AnderssonRendall}. Here we find this to be
the case only in the limit of small data $\hat U_\epsilon(\tau)$.

As this class of continuations seems to be locally CSS near the
singularity, it is interesting to study the exactly CSS solutions from
the point of view of a DSS ansatz. Starting from a generic DSS scalar
field (\ref{generalphi}) we introduce a new (gauge-dependent)
variable
\begin{equation}
\label{Wequation}
W \equiv U +  \frac{f}{x} (U+V) = \sqrt{2\pi G}(\kappa + \psi_{,\tau}),
\end{equation}
which coincides with $-V$ at the future lightcone $f/x=-1$, and obeys
the equation
\begin{equation}
xW_{,x}=-U_{,\tau} .
\end{equation}
In exactly CSS solutions of the system the metric functions and $U,V$
are independent of $\tau$, and hence $W$ is constant with value
$W=\sqrt{2\pi G}\kappa$. The CSS solution with $\kappa=0$ was
found in closed form by Roberts \cite{Roberts} and is described in
Appendix~\ref{appendix:roberts}. The CSS solutions with
$\kappa\ne 0$ were studied numerically by Brady \cite{Brady}.  

We find that the small $\hat U_\epsilon(\tau)$ continuations locally
approach a Roberts solution (with $\kappa=0$) with a value of the
parameter $p$ of the Roberts solution that depends on $\tau$, and not,
as one might have expected, one of Brady's solutions. This is discussed
in Appendix~\ref{appendix:roberts}.

\section{Global images of the Choptuik spacetime}
\label{section:globalimages}


Figures \ref{3dinnerplots}, \ref{choptuon_outer},
\ref{choptuon_future} and \ref{choptuon_singular} show in a very
detailed way the structure of the Choptuik spacetime. However it is
difficult to get from them an idea of what it looks like globally.
In this final Section before the conclusions we present a number of
additional figures that will fill this gap. We do it in two very
different ways.


\subsection{Global coordinate systems}


As shown in Fig.~\ref{figure:preCH}, our three $(\tau,x)$ patches match
continuously, but we do not expect the resulting global coordinate system
to be differentiable at the interfaces between the patches. 
The critical spacetime itself, however, is differentiable 
(analytic at the past
lightcone and $C^1$ at the CH), and it must be possible to construct
global coordinate systems which are at least $C^2$.

\begin{figure*}
\includegraphics[width=12cm]{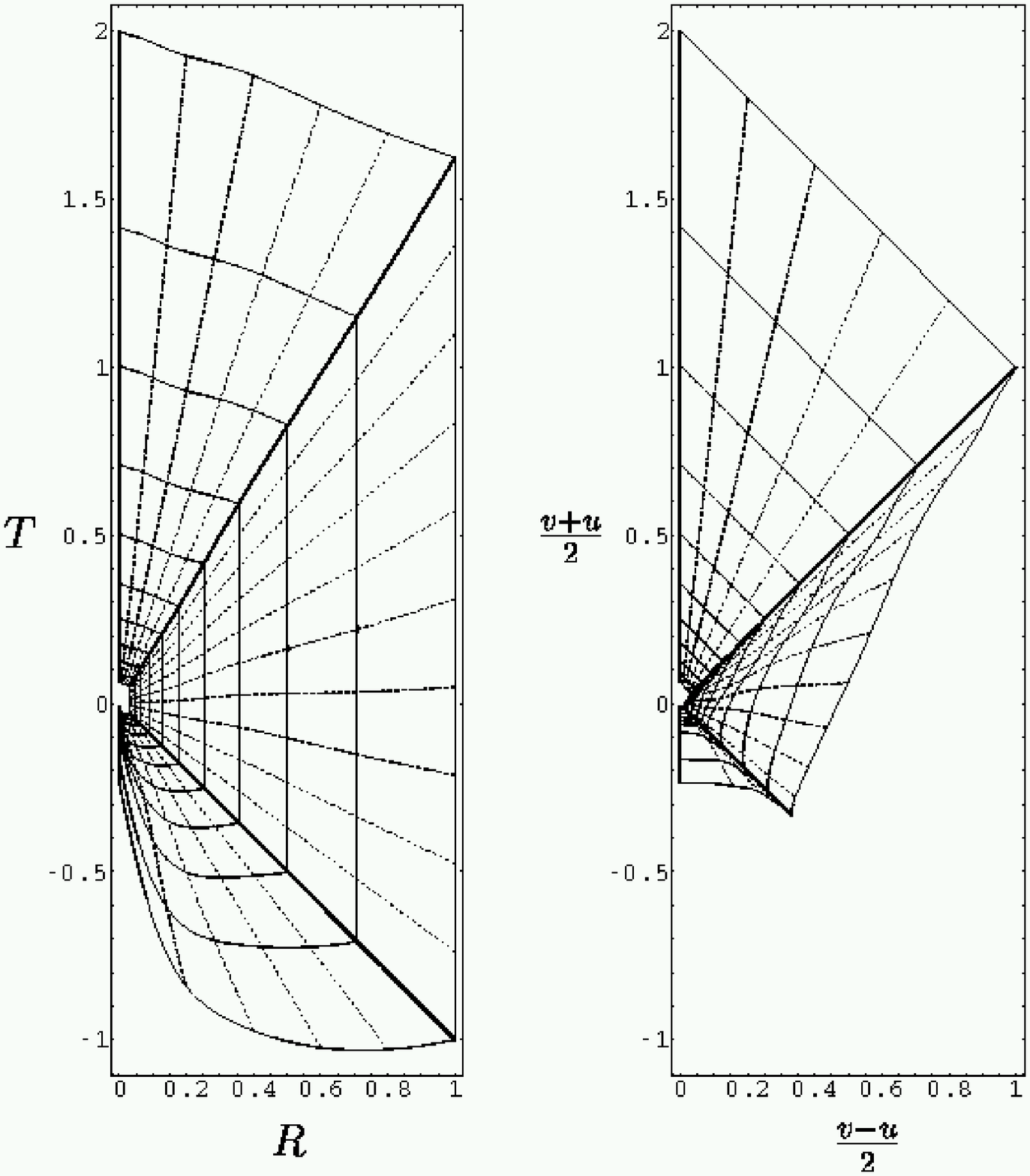}
\caption{ \label{coordinates}
Spacetime diagrams for a single period of the Choptuik spacetime in
differentiable coordinate systems (synchronous-area on the left and
double null on the right). It is possible to construct the whole
spacetime by adding rescaled (by a factor $e^\Delta=31.357$ or its
inverse) blocks to the center and outside the figures.
Continuous lines represent $\tau=const.$
lines and dashed lines represent $x=const.$ lines. By our gauge choices
times $T$ and $(v+u)/2$ coincide on the centre worldline.}
\end{figure*}

One simple possibility is synchronous slicing plus area gauge:
\begin{equation}
ds^2= -dT^2 + 2B(T,R)dTdR + C(T,R) dR^2 + R^2 d\Omega^2.
\end{equation}
We add the gauge condition $T=-R$ on the past lightcone of the
singularity. The coordinate transformation $T(\tau,x), R(\tau,x)$ can
be easily integrated and it is shown on the left panel of
Fig.~\ref{coordinates} for a single $\Delta$-period in $\tau$.

Another simple possibility is double null coordinates:
\begin{equation}
ds^2= -\omega(u,v)du dv + R^2(u,v)d\Omega^2
\end{equation}
with gauge condition $u=v=T$ at the center, where $T$ is the
time coordinate constructed in the previous paragraph. The coordinate
change $u(\tau,x), v(\tau,x)$ is shown on the right panel of
Fig.~\ref{coordinates} for a single $\Delta$-period in $\tau$.

\begin{figure*}
\includegraphics[width=15cm]{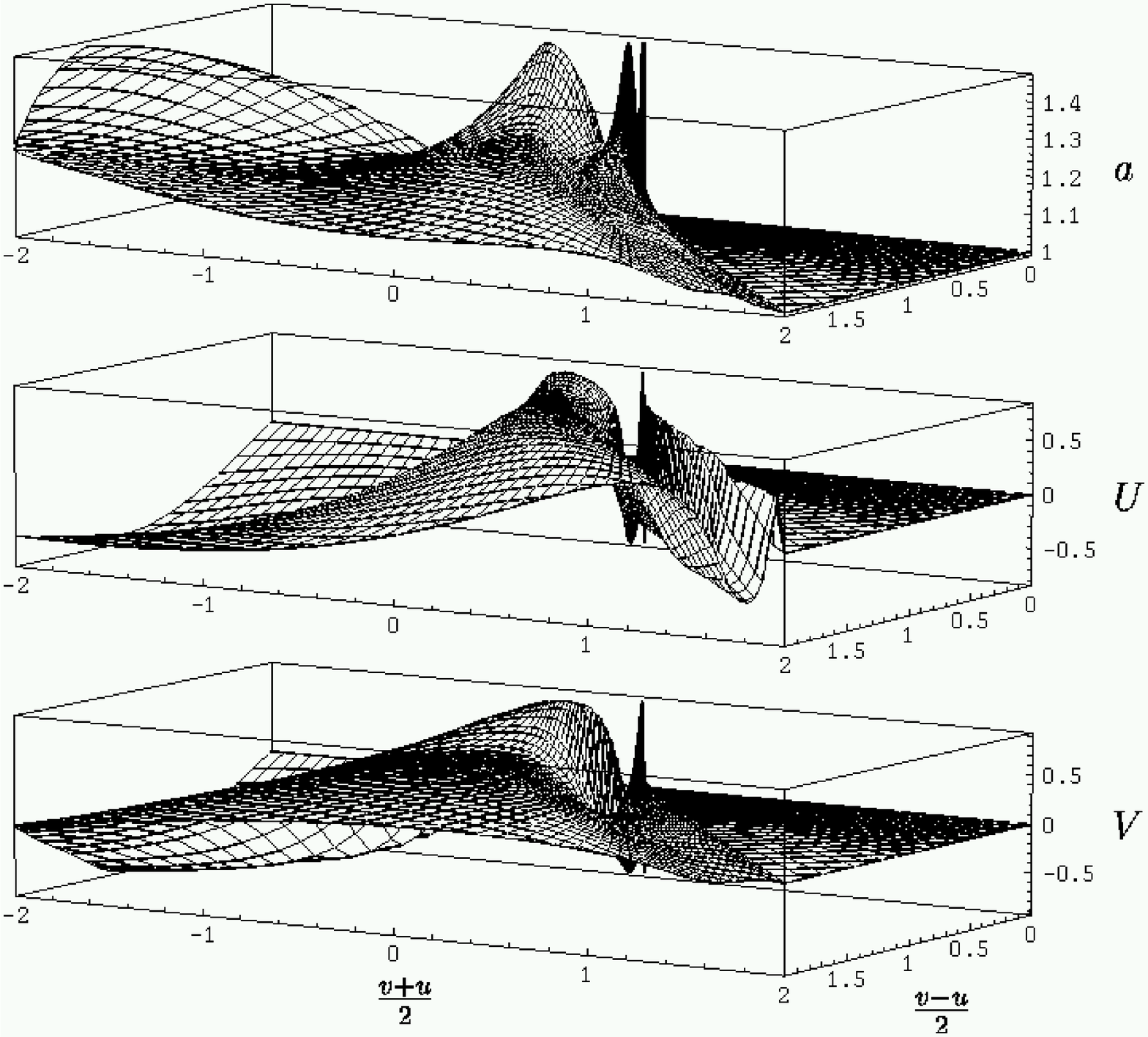}
\caption{ \label{fieldsuv}
Scalars $a$, $U$ and $V$ as functions of the double null variables
$u$ and $v$. Grid lines are lines of constant $x$ or $\tau$. Time is
increasing from left to right, and the central world line is at the back.
}
\end{figure*}

The fields $a$, $U$ and $V$ in double
null coordinates are shown in Fig.~\ref{fieldsuv}. As these fields 
are spacetime scalars, they should
be analytic in these coordinates at the past lightcone, and $C^1$ at
the Cauchy horizon. In the plot it 
looks as if they are only $C^0$ at the Cauchy
horizon, but that is due to a lack of resolution in the plot: the
slopes on the two sides of the Cauchy horizon are dominated by
$a_{1+\epsilon}$ and $a_{1+2\epsilon}$, which are discontinuous at the
resolution of the plot, but close enough to the Cauchy horizon, the
slope becomes $a_1$, which is continuous.

\subsection{Dynamical phase space portraits}

\begin{figure*}
\includegraphics[width=15cm]{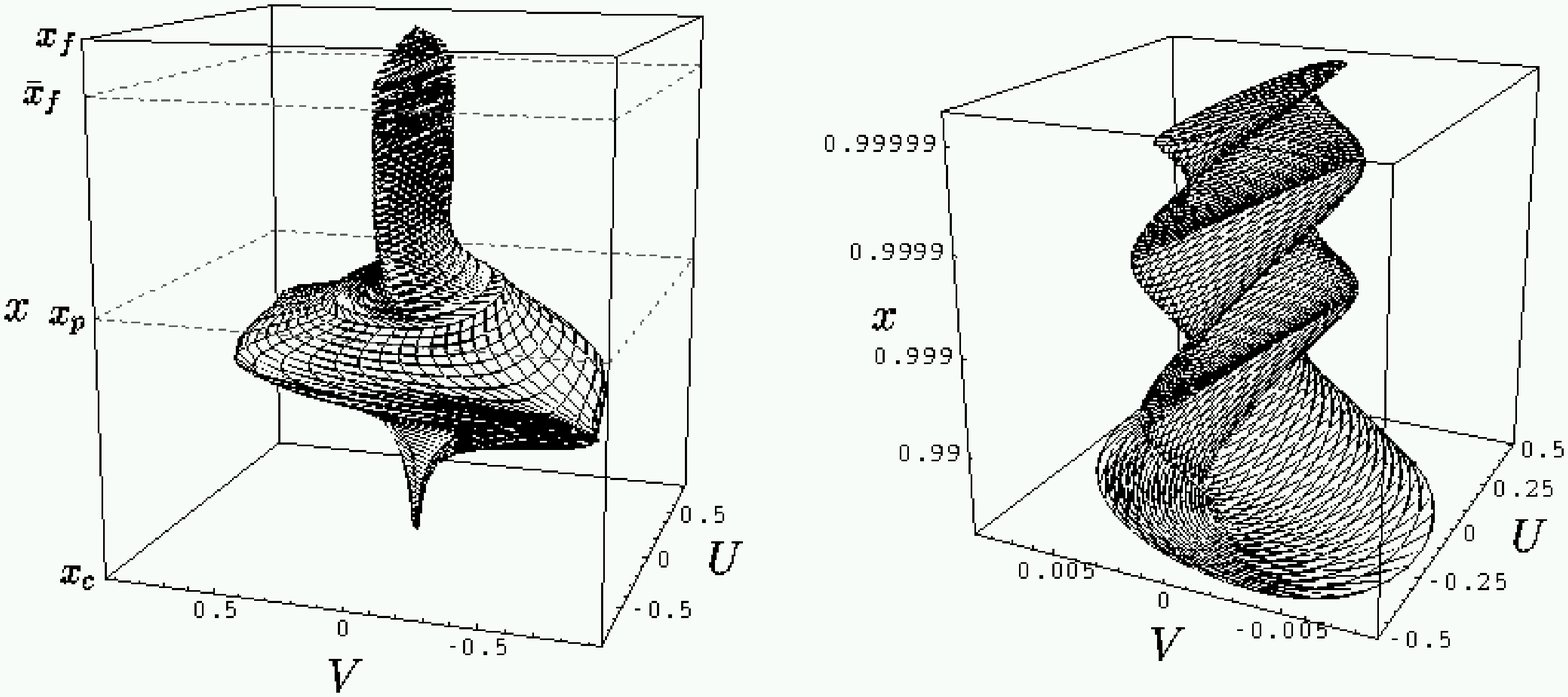}
\caption{ \label{dynbefore}
Left: Phase-portrait of the Choptuik spacetime from the regular past
centre at $x_c$ to the CH at $x_f$. The $x$-axis has been highly
distorted in order to show all interesting details: the axis is
logarithmic between $x_c$ and $x_p$ with accumulation point at $x_c$
(note that the label $x_c$ has been situated at finite distance for
convenience); the axis is logarithmic with accumulation at $x_f$ between
$x_p$ and $\bar{x}_f$; the axis has been transformed from $x$ to 
$x_f-x^{1/\epsilon}$ to show the decay of the function $U$ towards 
$x_f$ (this is a semi-analytical extrapolation of our numerical data).
Vertical generatrices are $\tau=const.$ lines.
Right: a reduction in the $V$ axis in the region between $x_p$ and $\bar{x}_f$
to show the oscillations. Compare with Fig.~\ref{Ulog}.
}
\end{figure*}
\begin{figure}
\includegraphics[width=\columnwidth]{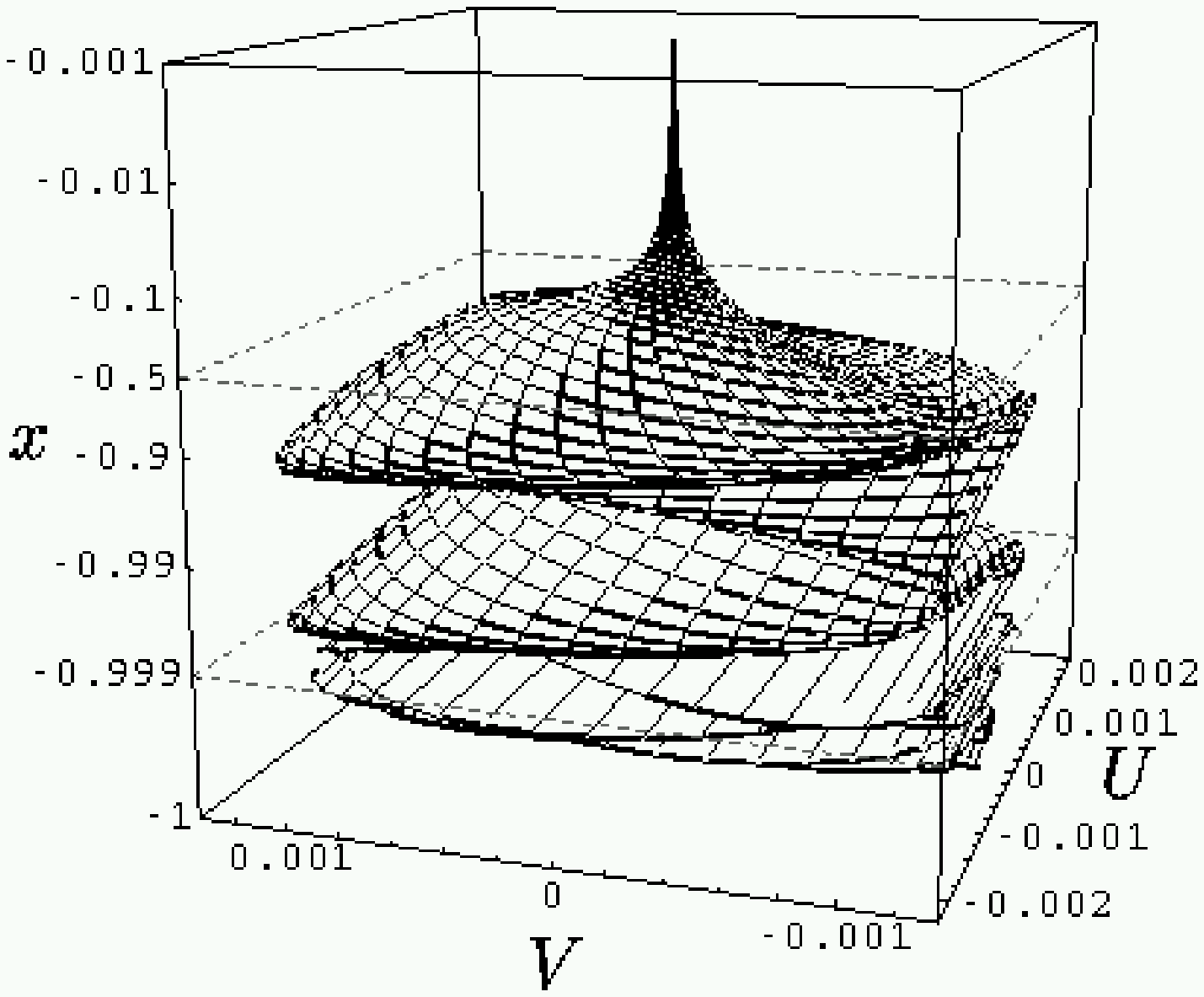}
\caption{ \label{dynafter}
Phase-portrait of the regular continuation of the Choptuik spacetime
from the CH at $x_f=-1$ to the regular center at $x_r=0$. Again, from
$x=-1$ to $x=-0.999$ we have distorted the axis to show the infinite
number of oscillations that pile up there. From $x=-0.999$ to $x=-0.5$
we use a logarithmic axis and from this middle point to $x=-0.001$ we
use a different logarithmic axis.
}
\end{figure}

We shall now consider the spherical DSS scalar field as an
infinite-dimensional dynamical system where $x$ is the ``time''
coordinate. The dynamical variables in this system are $U(\tau)$,
$V(\tau)$ and $f(\tau)$ (or $b(\tau)$ in the outer patch).  The
variable $a(\tau)$ is not independent, but given by a constraint.
However, many solutions of the dynamical system correspond to the same
spacetime, namely all that are related by the coordinate
transformations (\ref{dssgaugefreedom}).
The pair of periodic
functions $U(\tau)$ and $V(\tau)$ describes a closed, possibly
self-intersecting curve in the $(U,V)$ plane. An entire evolution in $x$
gives a surface that is topologically a cylinder, which we may call a
phase portrait of the solution. Clearly the surface itself is
invariant under the coordinate change $\tau\to\tau+\psi(x,\tau)$. If
we consider $\tau$ as our ``space'' coordinate in the usual ``3+1''
split, then by looking at the phase portrait we have eliminated the
``spatial'' gauge freedom. The ``slicing'' freedom
$x\to\varphi(x,\tau)$ however does change the shape of the phase
portrait, so it is not completely gauge-invariant. 
The Choptuik spacetime up to the CH is given in Fig.~\ref{dynbefore}
and its regular continuation is shown in Fig.~\ref{dynafter}.

Imposing CSS means that the system is independent of $\tau$ and hence
the phase-portrait in $(U,V,x)$ reduces to a line that can be easily
projected on the $(U,V)$ plane. Then the whole evolution of the system
can be described by this curve in the $(U,V)$ plane, which is now
completely gauge-invariant. This is essentially what Brady has done
\cite{Brady}.

\begin{figure}
\includegraphics[width=\columnwidth]{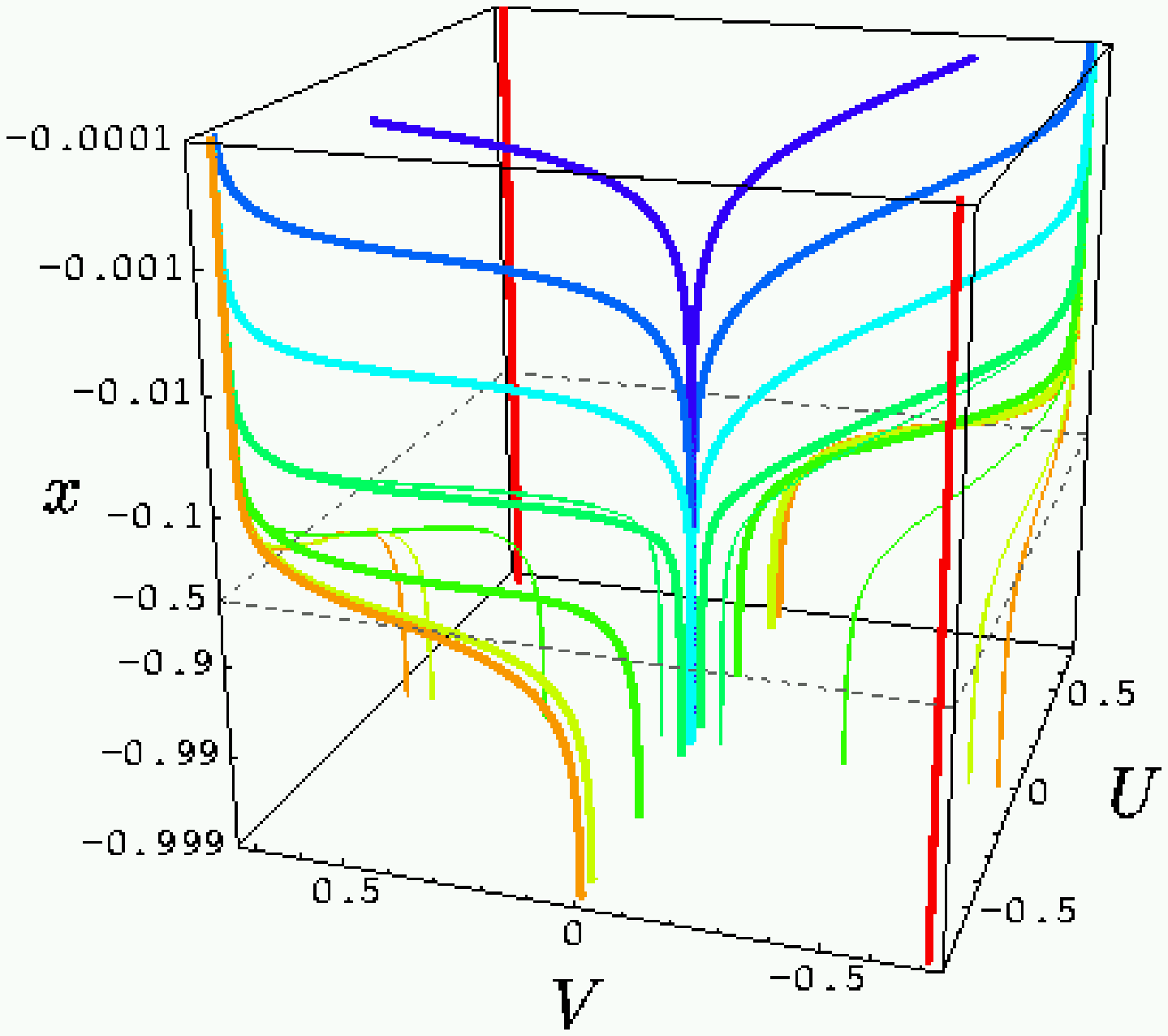}
\caption{ \label{Roberts}
Phase-lines of the Roberts spacetime. The colors encode different
values of $p$, for both branches: 1 (red), 0.99 (orange), 0.9 (yellow),
0.5 (light green), 0.1 (dark green), 0.01 (light blue), 0.001 (dark
blue) and 0.0001 (purple). The future (past) branch of the singularity
is denoted with thick (thin) lines starting with $V=0$ ($U=0$). Note 
that the vertical lines $U=V=\pm 1/\sqrt{2}$ corresponding to $p=1$
are unstable. Note also that the smaller $p$ is, the longer the curves
stay close to the unstable flat line $U=V=0$.
}
\end{figure}

\begin{figure}
\includegraphics[width=\columnwidth]{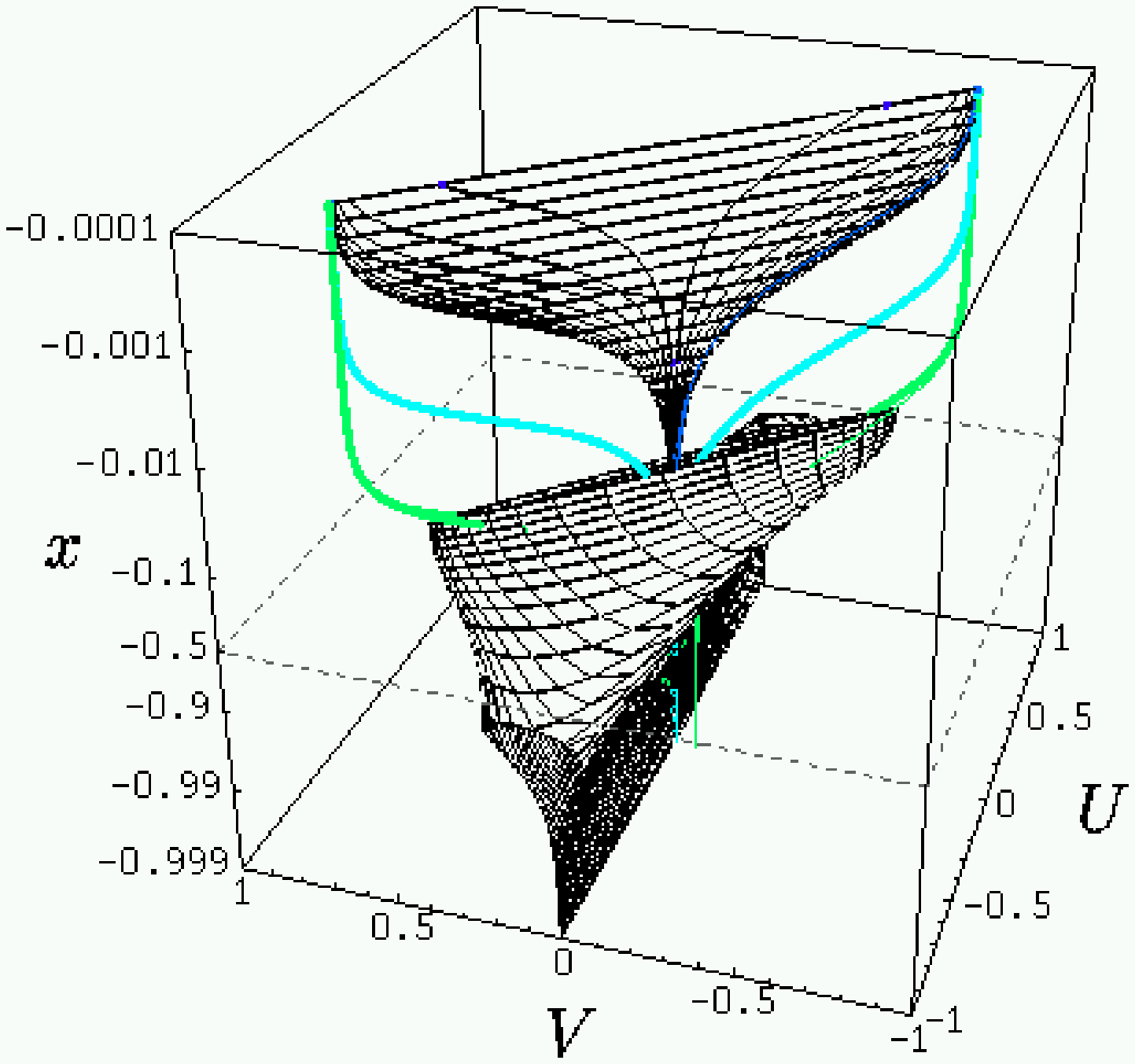}
\caption{ \label{dynsingular} Phase-portrait of two singular
continuation of the Choptuik spacetime (upper: initial $\hat
U_\epsilon$ of order $10^{-3}$; lower: initial $\hat U_\epsilon$ of
order $10$), together with a few Roberts solutions for
$p=0.1,0.01,0.001$. The $p=0.001$ solution almost coincide with the
external envelope of the upper phase-portrait and sets a maximum for
the values that are realized in this solution. The lower
phase-portrait is shown only for $-0.999<x\le-0.4$. For $-0.4<x<0$ it
is essentially on the diagonal in the $(U,V)$ plane although the
amplitude of the oscillations is not the same as in the Roberts
solution. Note that for small $x$ this phase portrait has been
truncated also in $U$, because $U$ becomes large. For $x\simeq -1$ the
phase portrait is essentially on the $U$-axis.}
\end{figure}

In order to understand the singular continuations of the critical
spacetime, we first look at phase portraits of the Roberts
spacetime. Fig.~\ref{Roberts} shows the phase-lines of the Roberts
solution for several values of $p$ in both branches (see also Appendix
\ref{appendix:roberts}). We see that the shape of the curves become
constant for very small values of $p$, but just translated along the
$\log x$-axis. On the other hand Fig.~\ref{dynsingular} gives the
phase-portrait two of our singular evolutions, for small and large
initial $\hat U_\epsilon$, respectively.  In both cases as
$U+V\to 0$ squeezes the phase cylinder into the diagonal,
following a Roberts solution in the former case, and not doing so in
the latter. This figure demonstrates why it is not possible to obtain
$a^2=2$ in the continuations of the Choptuik solution. The Roberts
spacetime with $p=1$ does have $a^2=2$ on the singularity, $p=0$
(Minkowski) has $a^2=1$, and all other $p$ have $a^2\to 0$. The $p=1$
Roberts solution has the phase line $U=V=\pm 1/\sqrt{2}$ line. To
reach $a^2=2$ approach a solution must approach this line from the
very beginning, because it is unstable. This requires $V$ of order
$1/\sqrt{2}$ at least and on the Cauchy horizon we only have $V$ of
order $10^{-3}$.

\section{Conclusions}
\label{section:conclusions}


In a companion paper \cite{spherical_dss} we investigate the
constraints that the kinematic assumptions of spherical symmetry and
discrete self-similarity impose on the causal structure.  The key
elements of our analysis were the self-similarity horizons -- radial
null geodesics that are mapped onto themselves by the self-similarity.
These come in two types: a ``fan'' connects a point singularity to a
piece of null infinity, while a ``splash'' connects a piece of null
singularity to a point at null infinity. All spherically symmetric and
self-similar spacetimes can be enumerated in terms of a sequence of
fans and splashes. We found that the singularity is central and
consists of a middle segment which is either a point or null line,
flanked by two segments which can be timelike, spacelike, or null
lines, or which can be absent. There are never two or more
disconnected singularities. 

In this paper, we have focussed on scalar field matter and in
particular Choptuik's critical solution in gravitational
collapse. This solution is found as an intermediate attractor in the
evolution of regular asymptotically flat scalar field initial data
near the black hole threshold \cite{Choptuik}. Independently, it can
be constructed as the solution of a nonlinear PDE boundary value
problem from the assumptions of spherical symmetry, discrete
self-similarity, and analyticity at the past center and past lightcone
of the singularity \cite{critscalar}. It can then be uniquely
continued up to the future lightcone (Cauchy horizon) of the
singularity.  Ref.~\cite{critscalar} found that the scalar field
oscillates roughly as $\cos(\ln|y|)$ as the Cauchy horizon $y=0$ is
approached. It was suggested on theoretical grounds that these
oscillations are damped by a factor $|y|^\epsilon$ with $\epsilon>0$,
but the numerical
value of $\epsilon$ was too small to distinguish it from zero
numerically.

We have repeated the numerical calculations of Ref.~\cite{critscalar}
from scratch and have increased the accuracy by more than four orders of
magnitude. We have found the positive value for $\epsilon$ given in
Eq.~(\ref{epsilon_value}). Therefore the scalar field is continuous on
the Cauchy horizon, with null data given in Eq.~(\ref{V_f_numerical}).
The structure of the fields near the Cauchy horizon is quite
complicated, and has been discussed in Subsections 
\ref{section:singularCH} and \ref{section:expansion}.

All possible DSS continuations beyond the Cauchy horizon are
determined by the (fixed) null data on the Cauchy horizon, and one
(arbitrary) periodic function $\hat U_\epsilon(\hat\tau)$ that can be
thought of as data emerging from the naked singularity. (In the
absence of the naked singularity, null data on the Cauchy horizon would
alone determine the continuation.) There is a unique DSS continuation
that has a regular timelike center except for a naked point singularity.
For all other values of the arbitrary function
$\hat U_\epsilon(\hat\tau)$, the continuation has a timelike central
singularity with infinite negative mass. In particular,
neither Fig.~\ref{figure:spacelikesingular}, nor the even
more exotic spacetime diagrams found in self-similar perfect fluid
solutions \cite{fluidcss} can arise as continuations of the Choptuik
solution (although they can probably arise in other DSS scalar field
solutions). 

The global structure of the Choptuik solution is of interest
partly because of the connection between critical collapse and cosmic
censorship. Choptuik's work has established that
(assuming the validity of general relativity at arbitrarily high
curvatures) a naked singularity can be formed in the spherical
Einstein-scalar field system from generic regular and asymptotically
flat initial data by fine-tuning any one parameter to the black hole
threshold. Solutions with a naked singularity therefore form a subset
of codimension one (the black hole threshold in the space of initial
data) in all solutions arising from regular data. The Choptuik
spacetime is an attractor on the critical surface and therefore,
assuming that it is actually a global attractor on the surface, we can
conclude that every naked singularity will have, at least in a
neighborhood of the singularity, the structure of the Choptuik
spacetime.

We can now summarize this structure as follows: The curvature at the
Cauchy horizon of the singularity is finite but not differentiable,
with an infinite number of damped oscillations piling up as the Cauchy
horizon is approached. The continuation of the spacetime beyond the
Cauchy horizon is not unique, but we have shown that if it is DSS then
the naked singularity is either a single point or timelike with
infinite negative mass. The spacetime near the timelike singularity is
locally CSS and velocity-dominated.

The question remains whether this structure is stable against
perturbations which break self-similarity and/or spherical
symmetry. Because the background is spherically symmetric and periodic
in $\tau$, all such perturbations are of the form
$e^{\lambda\tau}f(x,\tau)$, where $f(x,\tau)$ is periodic, times a
suitable scalar, vector or tensor spherical harmonic. We have shown in
the past \cite{critscalar,critscalar2} that all but one of these modes
decay, with ${\rm Re}\lambda<0$. (The one growing mode determines the
critical exponent for the black hole mass.)
The functions $f(x,\tau)$ were calculated explicitly
only between the regular center and past lightcone, but as they are by
construction analytic at the past lightcone, they can be analytically
continued up to the Cauchy horizon, with the same $\lambda$. 

What needs to be done is to check that the functions $f(x,\tau)$
remain bounded as $x$ approaches the Cauchy horizon. We expect that
this is true. Nolan and Waters \cite{NolanWaters} have investigated a
massless minimally coupled, non-spherically symmetric scalar field
propagating on a class of spherically symmetric CSS spacetimes, and
find that its gradient remains finite at the Cauchy horizon. The
structure of the full perturbation equations is similar.

As the functions $f(x,\tau)$ cannot be analytic at the Cauchy horizon
(because the background is not), the spectrum of $\lambda$ need not be
the same in the continuation beyond the Cauchy horizon, and is likely
to depend on the choice of continuation.

We have finally given, for the first time, a number of global
images of the Choptuik spacetime, trying to minimize the gauge content
of the pictures.


\acknowledgments
We would like to thank James Vickers for helpful discussions. This
research was supported by EPSRC grant GR/N10172/01. 


\appendix



\section{Review of coordinate systems for given $F$} 


In this Appendix we review a variety of coordinate systems defined by
giving $F$ as a function of $\tau$ and $x$. This class covers the
coordinate systems used in
Refs. \cite{OriPiran,critscalar,Brady,Garfinkle2+1,GoliathNilssonUggla},
but does not include the Lagrangian fluid coordinates used in
\cite{CarrColey}. We classify the gauges within this class in three
stages. 1) We fix $F(\tau,x)$. 2) We impose an algebraic relation
between $A$, $B$ and $C$. 3) We parameterize $A$, $B$ and $C$ in terms
of two functions, say $a$ and $f$. (It will turn out to be useful to
always use the scalar $a$ defined above as one of the two parameters.)

In a DSS spacetime, the unknowns are periodic in $\tau$. We therefore
solve the Einstein equations by ``evolving'' in $x$, with periodic
boundary conditions in $\tau$. When we use the Einstein equations to
eliminate the metric derivative $P$, $U$ and $V$ obey a pair of
transport equation of the form
\begin{equation}
U_{,x}=(\dots)\,U_{,\tau}+(\dots),
\quad V_{,x}=(\dots)\,V_{,\tau}+(\dots),
\end{equation}
where the dots stand for a known function of $U$, $V$, $a$ and $f$ but
not their derivatives.  By a suitable parameterization of $A$, $B$ and
$C$ in terms of $a$ and $f$, the Einstein equations can always be
brought into the form of two ODEs in $x$, which for our purposes are
evolution equations, and one ODE in $\tau$, which for our purposes is
a constraint:
\begin{equation}
\label{standardform}
f_{,x}=(\dots), \quad
a_{,x}=(\dots), \quad
a_{,\tau}=(\dots).  
\end{equation}
This last equation can be made linear in $a^{-2}$ (that is, in $\mu$)
by a suitable choice of $f$, at least in all three coordinate systems
we use. This inhomogeneous linear equation can then be solved uniquely
for $a$ in terms of $U$, $V$ and $f$, using the fact that $U$, $V$ and
$f$ are periodic in $\tau$ and that we require $a$ to be periodic,
too. In a CSS solution, where nothing depends on $\tau$, the
$a_{,\tau}$ constraint becomes an algebraic equation linking $U$, $V$,
$a$ and $f$.


\subsection{Past patches with $F=x$}


The regular center is both timelike and an $x$ line, but $\tau$ is
finite there. This requires $F=0$ at the center, and $F>0$
elsewhere. The simplest choice is $F=x$. On a past patch, it is
possible to choose the $\tau$ lines to be timelike, null or spacelike,
or to change signature. We consider only the first two possibilities

\paragraph{Bondi past patch} 

If we choose the $\tau$-lines to be null everywhere, this means
$C=0$. We then have $B<0$ between the past center and the past light
cone, and $B>0$ beyond the past lightcone, so that $A$ and $B$ both
change sign at the past lightcone. $F=x$ and $C=0$ has been used by
Brady \cite{Brady} to investigate CSS solutions with scalar field
matter. He used the Bondi metric coefficients that are traditionally
called $g$ and $\bar g$ as parameters. In a DSS solution this gives a
constraint for a combination of $g$ and $\bar g$, as well as evolution
equations for $g$ and $\bar g$. The best parameterization we have
found uses the mass function $a$ and the Bondi metric coefficient $f=g/a^2$:
\begin{eqnarray}
A&=&-g(\bar g\mp 2x)=- a^2 f(f\mp 2x) , \\
B&=&\mp g=\mp a^2 f, \\
C&=&0, \\
F&=&x,
\end{eqnarray}
where the upper sign applies when $\tau$ is an outgoing null
coordinate, and the lower sign applies when $\tau$ is an ingoing null
coordinate, assuming that $f>0$. This gives an ODE evolution equation
and a linear ODE constraint for $a^{-2}$, and an ODE evolution
equation for $f$.

\paragraph{Schwarzschild past patch}

If the $\tau$-lines are spacelike everywhere, this implies $C>1$. A
natural algebraic condition to impose is to make the $\tau$ lines
orthogonal to the $r$ lines. This means
$B(\tau,x)=-xC(\tau,x)$. $\tau$ is then $-\ln(-t)$ where $t$ is the
Schwarzschild time coordinate. A possible parameterization is in terms
of the metric coefficients often called $a$ and $\alpha$ in
Schwarzschild coordinates. ($a$ is the scalar we have defined
above). The best parameterization we have found is by $a$ and
$f=\alpha/a$, and is given in (\ref{pastpatch}) above.  It gives ODE
evolution equations for $a$ and $f$, and a linear ODE constraint
equation for $a^{-2}$. The transport equations for $U$ and $V$ are linear
in $U$ and $V$ because $P$ reduces to a function of $a$, $\tau$ and
$x$. The lightcones are at $f\pm x=0$ (although of course only one
lightcone can be covered at one time by a past patch).


\subsection{Outer patches with $F=1$}


On an outer patch that stretches from the past to the future lightcone
the $\tau$ lines must be timelike at least somewhere, and it is
possible to make them timelike everywhere. The simplest choice for $F$
compatible with all these possibilities is $F$=1.

\paragraph{Schwarzschild outer patch}

Assuming that the $\tau$ lines are timelike everywhere means $C<0$. We
have $B<0$ on the past lightcone and $B>0$ on the future lightcone,
so that $B$ has to change sign somewhere between. This ``$B$-surface''
is where the $\tau$ and $x$ lines are orthogonal. 
In \cite{critscalar} one of us used a coordinate system in
the outer patch that was based on Schwarzschild coordinates. As in the
past patch based on Schwarzschild coordinates, this meant
$B=-xC$. However, when we impose $B=-xC$ together with $F=1$, we face
an unexpected coordinate singularity at $x=0$. In particular, if we
again use $a$ and $f=\alpha/a$ to parameterize the metric, we obtain
an ODE constraint and an ODE evolution equation for $a$ as before on
the past patch, but for $f$ we obtain an equation of the form
\begin{equation}
f_{,x}={(\dots)f_{,\tau}+(\dots)\over x}
\end{equation}
In order to make the coordinates regular at $x=0$, we need to impose
the vanishing of the numerator of this equation there, and that is
effectively what was done in \cite{critscalar}. But this introduces an
additional boundary condition just to keep the coordinate system
regular on a surface where the solution itself is perfectly regular,
and that is why we do not use these coordinates here.

\paragraph{Best buy outer patch}

A better algebraic condition to combine with $F=1$ is $C=-1$. A
workable parameterization is in terms of the mass function $a$ and $B$
itself. This gives ODE evolution equations for $a$ and $B$, and a
(nonlinear) ODE constraint for $a$. The lightcones are at $a\pm
B=0$. If we replace $B$ by $b=B/a$, we have the parameterization given
above in (\ref{outerpatch}), and the ODE constraint for $a$ becomes
linear in $a^{-2}$, while the wave equation becomes linear in $U$ and
$V$.

\paragraph{$A=-C$: unworkable}

It is compatible with $F=1$ to assume that the $\tau$ lines ``bend
round'' to become spacelike at large and small $x$. 
means This that $C$ changes
sign. A natural choice is that $A$ and $C$ change sign together, and
we can impose this by setting $A=-C$. We have not found a way of
parameterizing this choice in a way that brings the Einstein equations
into the standard form (\ref{standardform}). Parameterizing with $B$
and $\mu$ gives two roots for $A$. Using $A$ and $\mu$ gives two roots
for $B$.

\paragraph{$x$ and $\tau$ orthogonal: unworkable}

We can also force the ``bending round'' by making $\tau$ and $x$ lines
orthogonal everywhere, or $B=0$. This implies that, while they
are independent, $A$ and $C$ change sign together. Parameterizing by
$A$ and $C$ we obtain an evolution equation and a constraint for $A$
and a constraint for $C$. But the constraint for $C$ is homogeneous,
$C_{,\tau}=N(A,U,V)C$, so that $C$ cannot be periodic. Therefore this
coordinate system is not compatible with CSS or DSS.

\paragraph{Forcing the lightcones: unworkable}

Instead of finding a working outer patch and then forcing the two
lightcones to fall on the lines $x=x_p$ and $x=x_f$ by imposing
$A(\tau,x_p)=A(\tau,x_f)=0$, we can make $A$ a given function of
$x$. The simplest such choice is $A=1-x^2$, which vanishes at $x=\pm
1$. (The two lightcones will be distinguished by the sign of $B$).
A parameterization that brings the Einstein equations into standard
form is 
\begin{equation}
A=1-x^2, \quad
B=bc, \quad
C=-c^2, \quad
F=1.
\end{equation}
This gives an evolution equation and a constraint for $b$ and a linear
constraint for $c$. However, the right-hand sides of all three
equations are of the form $N/b$. Furthermore, at $b=0$ the numerator
of $b_{,\tau}$ depends on the matter through $U^2+V^2$, while the
numerator of $b_{,x}$ depends on the matter through $U^2-V^2$. (The
numerator of $c_{,\tau}$ is proportional to that of $b_{,\tau}$.) That
means that we would have to impose two independent regularity
conditions at $b=0$. Together these would fix $U$ and $V$
completely. Therefore this coordinate system is not sufficiently
generic near the line $B=0$.


\section{Singular expansion around the Cauchy horizon}
\label{singular_expansion}


From (\ref{expansion1},\ref{expansion2}) we see that
\begin{equation}
D= y D_1(\tau)+y^2 D_2(\tau)+O\left(|y|^{2+\epsilon}\right), 
\end{equation}
The regular coefficients to $O(y)$ are as follows: $U_1(\tau)$ is the
unique periodic solution of
\begin{equation}
U_1'+(1-a_0^2-D_1)U_1+V_1-2a_0a_1U_0=0, 
\end{equation}
and
\begin{eqnarray}
V_1&=&-{\xi\over 2a_0}\left[(1-a_0^2)V_0+U_0+V_0'\right],\\
a_1&=&-{\xi\over 2}(U_0^2-V_0^2), \\
b_1&=&-{\xi'\over a_0}-{\xi\over 2a_0}\left(-3+a_0^2+2V_0^2\right),\\
b_2&=&-{a_1\over 2 a_0}b_1 \nonumber \\
&&-{\xi\over 4a_0}
\left(2a_0a_1-b_1U_0^2+b_1V_0^2+4V_0V_1\right), \\
D_1&=&-{a_0 b_1\over \xi} .
\end{eqnarray}
The regular coefficients to higher orders continue in this style, with
$U_n(\tau)$ the solution of a linear inhomogeneous ODE, while $V_n$,
$a_n$ and $b_n$ are given algebraically.

The nonvanishing singular coefficients up to $O(y^{1+3\epsilon})$ are
\begin{eqnarray}
a_{1+\epsilon}(\tau,x)&=&
\check a_{1+\epsilon}(\tau) \hat a_{1+\epsilon}(\hat\tau), \\
a_{1+2\epsilon}(\tau,x)&=&
\check a_{1+2\epsilon}(\tau) \hat a_{1+2\epsilon}(\hat\tau), \\
V_{1+\epsilon}(\tau,x)&=&
\check V_{1+\epsilon}(\tau) \hat V_{1+\epsilon}(\hat\tau), \\
U_{1+\epsilon}(\tau,x)&=&\sum_{i=1}^3
\check U_{1+\epsilon}^{(i)}(\tau) \hat U_{1+\epsilon}^{(i)}(\hat\tau),
\\
U_{1+2\epsilon}(\tau,x)&=&\sum_{i=1}^7 
\check U_{1+2\epsilon}^{(i)}(\tau) \hat U_{1+2\epsilon}^{(i)}(\hat\tau),
\\
U_{1+3\epsilon}(\tau,x)&=&\sum_{i=1}^6 
\check U_{1+3\epsilon}^{(i)}(\tau) \hat U_{1+3\epsilon}^{(i)}(\hat\tau).
\end{eqnarray}

We only give two examples of how these coefficients
are derived. Substituting the ansatz into the $a_{,x}$ equation and
isolating the terms of $O(y^\epsilon)$ in the result, we obtain
\begin{eqnarray}
(1+\epsilon)\check a_{1+\epsilon}(\tau) \hat a_{1+\epsilon}(\hat\tau)
+K \check a_{1+\epsilon}(\tau) \hat a_{1+\epsilon}'(\hat\tau)
= \nonumber \\
-\xi(\tau) U_0(\tau) \check U_\epsilon(\tau)\hat U_\epsilon(\hat\tau).
\end{eqnarray}
We solve this for all $\tau$ and $\hat\tau$ by setting
\begin{equation}
\check a_{1+\epsilon}(\tau)={\xi(\tau) U_0(\tau) \check
U_\epsilon(\tau)\over K}
\end{equation}
and by making $\hat a_{1+\epsilon}(\hat\tau)$ the unique solution of
the ODE
\begin{equation}
\hat a_{1+\epsilon}'+{1+\epsilon\over K} \hat a_{1+\epsilon} + \hat
U_\epsilon = 0.
\end{equation}

If we substitute the ansatz into the $U_{,x}$ equation 
(\ref{U_singular_eq}) and isolate the terms of $O(y^{1+\epsilon})$,
we find
\begin{widetext}
\begin{eqnarray}
D_1\left((1+\epsilon)\sum_i \check U_{1+\epsilon}^{(i)} 
\hat U_{1+\epsilon}^{(i)}
+K \sum_i U_{1+\epsilon} \hat U_{1+\epsilon}'^{(i)}\right)
+D_2\left((1+\epsilon)\check U_{\epsilon} \hat U_{\epsilon}
+K U_{\epsilon} \hat U_{\epsilon}'\right) 
\nonumber \\
=(1-a_0^2)\sum_i U_{1+\epsilon}^{(i)} \hat U_{1+\epsilon}^{(i)}
-2a_0a_1 \check U_{\epsilon} \hat U_{\epsilon}
-2a_0 \check a_{1+\epsilon} \hat a_{1+\epsilon} U_0
+\check V_{1+\epsilon} \hat V_{1+\epsilon}
+\sum_i \check U_{1+\epsilon}'^{(i)} \hat
U_{1+\epsilon}^{(i)}+(1+H')\sum_i 
\check
U_{1+\epsilon}^{(i)} \hat U_{1+\epsilon}'^{(i)} 
\end{eqnarray}
Because $KD_1=1+H'$, the derivatives $\hat U_{1+\epsilon}'^{(i)}$
cancel out. Taking also into account that $\hat V_{1+\epsilon}=\hat
a_{1+\epsilon}$ (this is an accident at this particular order), the
equation can be rewritten as
\begin{eqnarray}
\label{U1pe}
\sum_i \left[\check U_{1+\epsilon}'^{(i)} 
+\left(1-a_0^2-(1+\epsilon)D_1\right)
\check U_{1+\epsilon}^{(i)} \right]\hat U_{1+\epsilon}^{(i)}  \nonumber \\
+\left[-\epsilon D_2-2a_0a_2)\check U_\epsilon\right] \hat U_\epsilon
+\left[-KD_2\check U_\epsilon\right] \hat U_\epsilon'
+\left[\check V_{1+\epsilon}-2a_0\check a_{1+\epsilon}U_0\right] 
\hat a_{1+\epsilon}=0.
\end{eqnarray}
\end{widetext}
The terms in square brackets all depend on $\tau$ and each multiplies
a different known function of $\hat\tau$. To solve this for all $\tau$
and $\hat\tau$, we assign one term $\hat U_{1+\epsilon}^{(i)}$ to each
function of $\hat\tau$:
\begin{equation}
\hat U_{1+\epsilon}^{(1)}=\hat U_\epsilon,
\quad
\hat U_{1+\epsilon}^{(2)}=\hat U_\epsilon',
\quad
\hat U_{1+\epsilon}^{(3)}=\hat a_{1+\epsilon}.
\end{equation}
The corresponding coefficients $\check U_{1+\epsilon}^{(i)}$ are the
unique solutions of the ODEs
\begin{equation}
\check U_{1+\epsilon}'^{(i)} 
+\left(1-a_0^2-(1+\epsilon)D_1\right)
\check U_{1+\epsilon}^{(i)} + S^{(i)}=0,
\end{equation}
where the source terms $S^{(i)}$ can be read off directly from
(\ref{U1pe}).  The calculation of the other terms proceeds in a
similar manner at all orders. Note that the functions of type $\check
f$, where $f$ stands for $V$, $a$ and $b$, are given algebraically and
$\hat f$ obeys an ODE. For $U$ it is the other way around.

The limit $V_0=0\Leftrightarrow\epsilon=0$ of our expansion
exists. The regular $O(y^n)$ terms in the series vanish identically,
and so do some of the $O(|y|^{n+k\epsilon}$ terms but not all of
them. $a_1$ and $a_{1+\epsilon}$ for example vanish because they are
proportional to $U_0$, but $a_{1+2\epsilon}$ is proportional to
$U_\epsilon^2$, and so encodes the curvature to $O(y)$.

In the limit $\epsilon=0$ the curvature components proportional to
$U^2$ are no longer continuous at the CH because they are now periodic
in $\hat\tau=\tau-\ln|y|$, but for the same reason they remain
bounded. The components proportional to $UV$ and $V^2$ are still
continuous even then because $V$ is $O(y)$. $a$ is also still
continuous, with $a=1$ on the Cauchy horizon. 


\section{The Roberts solution}
\label{appendix:roberts}


In the notation of \cite{Oshiro}, the Roberts solution \cite{Roberts}
is given by \begin{eqnarray}
\label{Roberts1}
ds^2 & = & - du\,dv + r^2(u,v) \,d\Omega^2, \\ 
\label{Roberts2}
r^2(u,v) & = & \frac{1}{4}\left[(1-p^2)v^2 - 2vu + u^2\right], \\
\label{Roberts3}
\phi(u,v) & = & \frac{1}{\sqrt{16\pi G}} \log \frac{(1-p)v-u}{(1+p)v-u},
\end{eqnarray}
with $p$ a constant parameter. $p=0$ is Minkowski spacetime with zero
scalar field, and without loss of generality $p\ge 0$. Only the
regions $r^2>0$ are physical and without loss of generality, we
consider the right side ($v-u>0$) of the spacetime.

For $p\ne 0$, the lines $u=(1\pm p)v$ are central curvature 
singularities: the mass function $\mu=-p^2 uv/(4r^2)$ diverges on them.
The line $u=(1+p)v$, $v<0$ is timelike and has (infinite) negative mass,
and for $v<0$ forms the past branch of the Roberts singularity. The line
$u=(1-p)v$ is timelike with negative mass for $p<1$, null with zero
mass for $p=1$, and spacelike with positive mass $p>1$. For $v>0$ it
forms the future segment of the singularity.

Some of our solutions become asymptotically CSS and have 
$\kappa\simeq 0$, therefore approaching the Roberts spacetime. We know
that the solutions develop a timelike, negative mass singularity, which
must correspond to one of the two branches of the Roberts singularity.
In this Appendix we study the issue of which of the two branches is
actually approached in our numerical evolutions and for what values 
of $p$.

Our scalars $U$ and $V$ are
\begin{equation}
U= -\frac{p}{\sqrt{2}} \, \frac{v}{v-u} , \qquad
V=  \frac{p}{\sqrt{2}} \, \frac{u}{v-u-p^2v} ,
\end{equation}
where both denominators are positive in the region of interest.
It is clear that $p$ is covariantly defined by $U$ and $V$ on the light
cones $u=0$ or $v=0$, respectively. However, at the singularities 
$u=(1\pm p)v$ the scalars take values $U=-V=\pm 1/\sqrt{2}$ for any 
$0\ne p\ne 1$ in the Roberts solution. If we use an expansion using a 
generic self-similar coordinate $y$ such that $u/v=F(y)$ with the
singularity at $y=0$ [that is, $F(0)=1\pm p$]:
\begin{eqnarray}
\sqrt{2}\,U(y)&=& \pm 1 - \frac{y}{p}F'(0) + O(y^2), \\
\sqrt{2}\,V(y)&=& \mp 1 + \frac{y}{p}\frac{1\mp p}{1\pm p}F'(0) 
+ O(y^2).
\end{eqnarray}
Therefore $p$ is covariantly given by the ratio of the rates at which
$U$ and $V$ approach their values at the singularity:
\begin{equation} \label{UVratio}
\lim_{x\to 0}{dU\over dV} = - \frac{1\pm p}{1\mp p}.
\end{equation} 
This determines both $p$ and which branch of the singularity we
approach locally: The upper sign applies for $|dU/dV|>1$ (past
branch, turned upside down) and the lower sign for $|dU/dV|<1$ 
(future branch of the singularity).

We can relate the coordinates $(u,v)$ of (\ref{Roberts1}-\ref{Roberts3})
to our coordinate $x$ through
\begin{equation}
\left(1-\frac{u}{v}\right)^2 = p^2+(1-p^2)x^2 .
\end{equation}
This expression applies only for negative $x$ and so only covers
the future branch of the singularity. We obtain the expressions for
the past branch exchanging the role of the functions $U$ and $V$.
That explains why exchanging $U$ and $V$ in Eq. (\ref{UVratio})
amounts to a change of branch. It also shows that we have to change
branch four times per $\Delta$-period as we move through the four
quadrants determined by the lines $U+V=0$ and $U-V=0$.




\begin{thebibliography}{}


\bibitem{Choptuik} M. W. Choptuik, Phys. Rev. Lett.
{\bf 70}, 9 (1993).

\bibitem{AbrahamsEvans} A. M. Abrahams and C. R. Evans, 
Phys. Rev. Lett. {\bf 70}, 2980 (1993).

\bibitem{critreview} C. Gundlach, Living Reviews in Relativity {\bf
2}, 4 (1999), published electronically at
http://www.livingreviews.org. For an updated version see C. Gundlach,
Critical phenomena in gravitational collapse, to be published in
Phys. Rep., preprint gr-qc/0210101.

\bibitem{spherical_dss} C. Gundlach and J. M. Mart\'\i n-Garc\'\i a,
Kinematics of discretely self-similar spherically symmetric
spacetimes, in preparation.

\bibitem{critscalar} C. Gundlach, Phys. Rev. D {\bf 55}, 695 (1997).

\bibitem{Nolan} B. C. Nolan, Class. Quant. Grav. {\bf 18}, 1651
(2001).

\bibitem{OriPiran} A. Ori and T. Piran, Phys. Rev. Lett. {\bf
59}, 2137 (1987).

\bibitem{fluidcss} B. J. Carr and C. Gundlach, Phys. Rev. D {\bf 67},
024035 (2003).

\bibitem{Iserles} A. Iserles, {\em Numerical Analysis of Differential
Equations}, Cambridge University Press, 1996.

\bibitem{AnderssonRendall} L. Andersson and A. D. Rendall,
Commun. Math. Phys. {\bf 218} 479 (2001).

\bibitem{Roberts} 
M. D. Roberts, { Gen. Rel. Grav.} {\bf 21}, 907-939
(1989).

\bibitem{Brady} P. Brady, Phys. Rev.  D {\bf 51}, 4168 (1995).

\bibitem{critscalar2} J. M. Mart\'\i n-Garc\'\i a and C. Gundlach,
Phys. Rev.  D {\bf 59}, 064031 (1999).

\bibitem{NolanWaters} B. C. Nolan and T. J. Waters, Phys. Rev. D {\bf
66}, 104012 (2002).

\bibitem{Garfinkle2+1} D. Garfinkle, Phys. Rev. D {\bf 63}, 044007 
(2001).

\bibitem{GoliathNilssonUggla} B. J. Carr, A. A. Coley, M. Goliath,
U. S. Nilsson, and C. Uggla, Class. Quant. Grav. {\bf 18}, 303 (2001).

\bibitem{CarrColey} B. J. Carr and A. A. Coley, Phys. Rev. D {\bf 62},
044023 (2000).

\bibitem{Oshiro} 
Y. Oshiro, K. Nakamura, and A. Tomimatsu,
{ Prog. Theor. Phys.} {\bf 91}, 1265 (1994).



\end{thebibliography}
\end{document}